\documentclass[journal]{IEEEtran}

\usepackage{graphicx}
\usepackage{amssymb,amsmath}
\usepackage{multirow}
\usepackage{tabularx}
\usepackage{slashbox}
\usepackage{color}
\usepackage{lettrine}
\usepackage{mdwlist}
\usepackage{cite}

\newcounter{Mycount}
\newcounter{Mycount2}
\newcounter{Mycount3}

\begin{document}

\title{Performance Analysis of SPAD-based OFDM}

\author{\vspace*{0.5cm} Yichen Li, Majid Safari, Robert Henderson and Harald Haas}

\maketitle

\begin{abstract}

In this paper, an analytical approach for the nonlinear distorted bit error rate performance of optical orthogonal frequency division multiplexing (O-OFDM) with single photon avalanche diode (SPAD) receivers is presented. Major distortion effects of passive quenching (PQ) and active quenching (AQ) SPAD receivers are analysed in this study. The performance analysis of DC-biased O-OFDM and asymmetrically clipped O-OFDM with PQ and AQ SPAD are derived. The comparison results show the maximum optical irradiance caused by the nonlinear distortion, which limits the transmission power and bit rate. The theoretical maximum bit rate of SPAD-based OFDM is found which is up to 1~Gbits/s. This approach supplies a closed-form analytical solution for designing an optimal SPAD-based system.

\end{abstract}


\section{Introduction}

\lettrine[lines=2]{\textbf{C}}{URRENTLY}, high speed light emitting diodes (LEDs) and laser diodes (LDs) are mainly used as transmitters in optical wireless communication (OWC) systems. With a single LED, a OWC system can achieve data rates exceeding 3 Gb/s~\cite{TD2013}. However, the incoherent light output of the transmitters means that information can only be encoded in the intensity level. As a consequence, only real-valued and positive signals can be used for data modulation. Thus, OWC systems are usually considered to be modulated as an intensity modulation and direct detection (IM/DD) system~\cite{kb9701}. Unipolar modulation schemes with single-carrier, such as on-off keying (OOK), pulse position modulation (PPM) and pulse amplitude modulation (PAM), can be used in conjunction with IM/DD systems~\cite{kb9701,mz0601,wbcb0501,lrbk0901}. In order to fully use the limited modulation bandwidth of the device and achieve high data rates, orthogonal frequency division multiplexing (OFDM) is applied in OWC systems by utilizing adaptive bit and power loading~\cite{TD2013}. Unlike OFDM in radio frequency, optical OFDM (O-OFDM) requires real valued signals, and these are generated by imposing Hermitian symmetry on the information frame before the inverse fast Fourier transform (IFFT) operation during the signal generation phase. However, this decreases the spectral efficiency by half. Diverse O-OFDM modulation schemes have been realized and applied in OWC, such as DC-biased optical OFDM (DCO-OFDM), asymmetrically clipped optical OFDM (ACO-OFDM), unipolar OFDM (U-OFDM) and non-DC-biased OFDM (NDC-OFDM)~\cite{SPADOFDM,meh1001,emhp0701,tsh1201,mhh1201}.

Typically, highly sensitive photodiodes (PDs), such as positive-intrinsic-negative (PIN) diodes and avalanche photo diodes (APDs), are applied as receivers in OWC. However, when the OWC system is applied in low optical power and long distance transmission, such as in a gas well downhole monitoring system~\cite{ysh1401} and data transmission over plastic optical fibres~\cite{SPADOC}, the number of photons reaching the receivers are significantly less than in standard indoor OWC links. In these scenarios, conventional PDs have unsatisfactory performance because the transimpedance amplifier (TIA) significantly reduces the sensitivity of the receiver and limits the signal-to-noise ratio (SNR). As a consequence, these low power signals are buried in noise. Hence, when compared with conventional PDs, single photon avalanche diodes (SPADs) would be more suitable receivers in these scenarios. The SPAD detector does not require a TIA and thus the output signal is not distorted by thermal noise. In addition, as SPADs can even detect a single photon, a bit of information-carried photons can be received accurately. Therefore, the SPAD receiver can perform at significantly higher sensitivity and optical power efficiency than conventional PDs. In previous work~\cite{SPADOFDM}, an O-OFDM system with a SPAD receiver was presented and compared with state-of-the-art PD-received based O-OFDM systems. When the transmission speed is 1~Mbits/s, SPAD-based OFDM enhanced the sensitivity by 30.5~dB over the PD-based system.

However, a SPAD receiver can only detect one photon within a device specific dead time which constrains the ability to recover a signal. In addition, since the output of the detector is a photon count value, there is a maximum number of photons that the system can detect. This limits the maximum tolerable optical irradiance which results in a receiver nonlinear distortion. This means that the transmission power and maximum bit rate of SPAD-based OFDM are limited by the structure and design of SPAD receivers~\cite{SPADOFDMnon,AlmerGlobe}. The analytical model of the nonlinear distortion effect in O-OFDM with conventional PDs has been derived in~\cite{dtv0001,dsh2013non,dsh1102}. As the nonlinear effects in the conventional O-OFDM system are mainly caused by transmitter properties and modulation schemes, the SPAD receiver nonlinear distortion effect has not yet been reported. This study provides a complete analytical procedure to find the exact bit error rate (BER) of the SPAD-based OFDM by considering the receiver nonlinear distortion and the conventional distortions. The analytical model of SPAD-based OFDM can be used to find the limitation threshold in the system and also the theoretical maximum bit rate. In addition, as the current SPAD array is designed for image processing~\cite{SPADfunda}, the designed parameters may not be suitable for OWC. Based on the analytical model, a reliable approach is presented for the design of the SPAD array with some optimal parameters which are suitable for current OWC systems.

\begin{figure*}[!t]
\begin{center}
\includegraphics[width=0.96\textwidth]{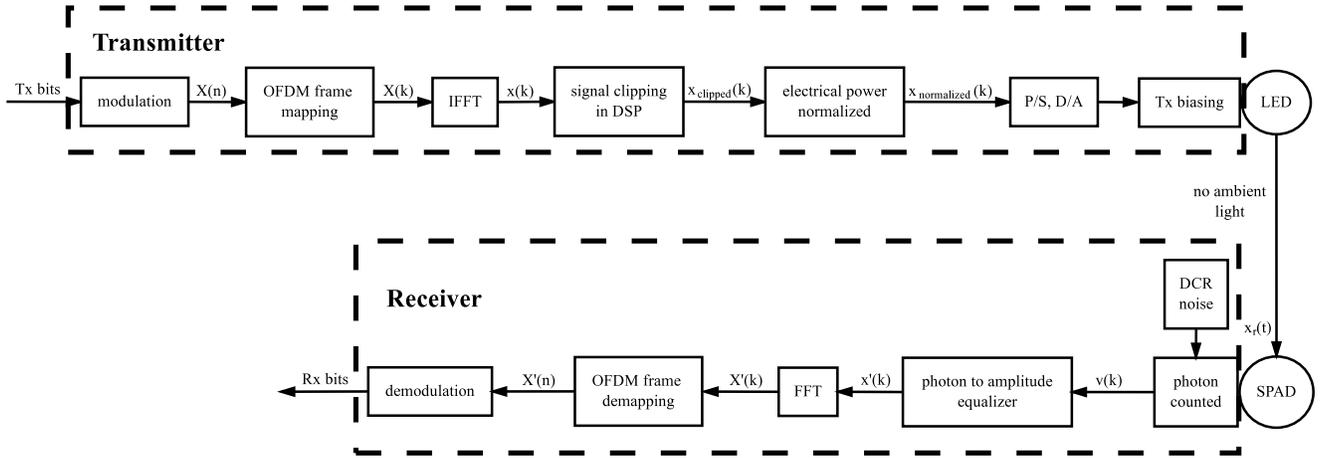}
\caption{Block diagram of the SPAD-based OFDM system.}
\end{center}
\end{figure*}

The rest of this paper is organized as follows. The system model of the SPAD-based OFDM system is described in Section~\uppercase\expandafter{\romannumeral2}. The nonlinear distortion in the SPAD receiver is presented in Section~\uppercase\expandafter{\romannumeral3}. The theoretical analysis of SPAD-based OFDM with nonlinear distortion is derived in Section~\uppercase\expandafter{\romannumeral4}. The numerical and analytical results of the system and discussion on the performance of the system are given in Section~\uppercase\expandafter{\romannumeral5}. Finally, conclusions are given in Section~\uppercase\expandafter{\romannumeral6}.

\section{SPAD-based OFDM}

The system model of OFDM with SPAD receivers is shown in Fig.~1.

\subsection{Optical OFDM Modulation}

At the transmitter, the input bit stream is transformed into complex symbols, $\textbf{X}(n)$, by a $M$-quadrature amplitude modulation (QAM) modulator, where $M$ is the constellation size. The symbols are allocated on to $N$ subcarriers, $\textbf{X}(k)$, $k = 0, \cdots,N-1$. In OFDM, $N$ denotes the size of IFFT/FFT, where $N$ is set to 2048. In general, two standard techniques, DCO-OFDM and ACO-OFDM, are used to obtain positive and real-valued OFDM symbols~\cite{dsh1102}. In DCO-OFDM, $N/2-1$ symbols in $\textbf{X}(n)$, $n = 1, \cdots,N/2-1$, are put into the first half of subcarriers and the DC subcarrier (the first subcarrier) is set to zero. In ACO-OFDM, $N/4$ QAM symbols in $\textbf{X}(n)$, $n = 1, \cdots,N/4$, are mapped on to half of the odd subcarriers of the OFDM frame, $\textbf{X}(k)$, $k = 1,3,5, \cdots, N/2-1$. At the same time, the even subcarriers are set to zero. In both ACO-OFDM and DCO-OFDM, Hermitian symmetry is applied to the rest of the OFDM frame in order to obtain real-valued symbols through the IFFT block. Since transmitters can only send unipolar signals, the real-valued OFDM symbols need to be clipped. In DCO-OFDM, a DC bias is added to make the signal unipolar~\cite{dsh1102}. In practice, the value of the DC bias, which is related to the average power of the OFDM symbols, is defined as:
\begin{equation}
B_{\rm DC} = \beta\sqrt{ {\rm E}\left[ \textbf{x}^2(k) \right] },
\label{Bdc}
\end{equation}
where ${\rm E}[.]$ represents the statistical expectation; $\textbf{x}(k)$ is the OFDM symbol frame vector; and $10\log_{10}(\beta^2+1)$ is defined as the bias level in dB. The bias level in the current simulations is set to 7 dB and 13 dB, which are adopted from~\cite{SPADOFDM} for consistency. After the DC bias, the OFDM frame is simply clipped by:
\begin{equation}
\textbf{x}_{\rm{clipped}}(k) = \left\{\begin{array}{ll} \textbf{x}_{\rm{biased}}(k), & \textbf{x}_{\rm{biased}}(k) \geq 0, \\ 0, & \textbf{x}_{\rm{biased}}(k) < 0, \end{array}\right.
\end{equation}
where $\textbf{x}_{\rm{biased}}(k)$ is the DC biased symbol which is calculated as $\textbf{x}_{\rm{biased}}(k) = \textbf{x}(k) + B_{\rm DC}$. The clipped unipolar symbol is denoted by $\textbf{x}_{\rm{clipped}}(k)$. In ACO-OFDM, since symbols are antisymmetric, clipped unipolar symbols are obtained by setting the negative part to zero. In the simulation, after being transformed into an optical intensity signal, the clipped signal is transmitted by the LED transmitter.

\subsection{SPAD Receiver}

A SPAD is an APD which is biased beyond reverse breakdown in the so called `Geiger' region. In this mode of operation, a SPAD triggers billions of electron-hole pair generations for each detected photon. In other words, in `Geiger' mode, a SPAD emits a very large current by receiving a single photon and thus can essentially be modelled as a single photon counter. The photodetection process of an ideal photon counter can be modelled using Poisson statistics which describe the shot noise effect (\cite{MS001} and references therein). 

In this study, in order to increase the capacity of the photon counts, an array of SPADs which outputs the superposition of the photon counts from the individual SPADs is considered~\cite{SPAD2013}. Some significant parameters of the SPAD array are introduced as follows.

\begin{figure}[!t]
\begin{center}
\includegraphics[width=0.48\textwidth]{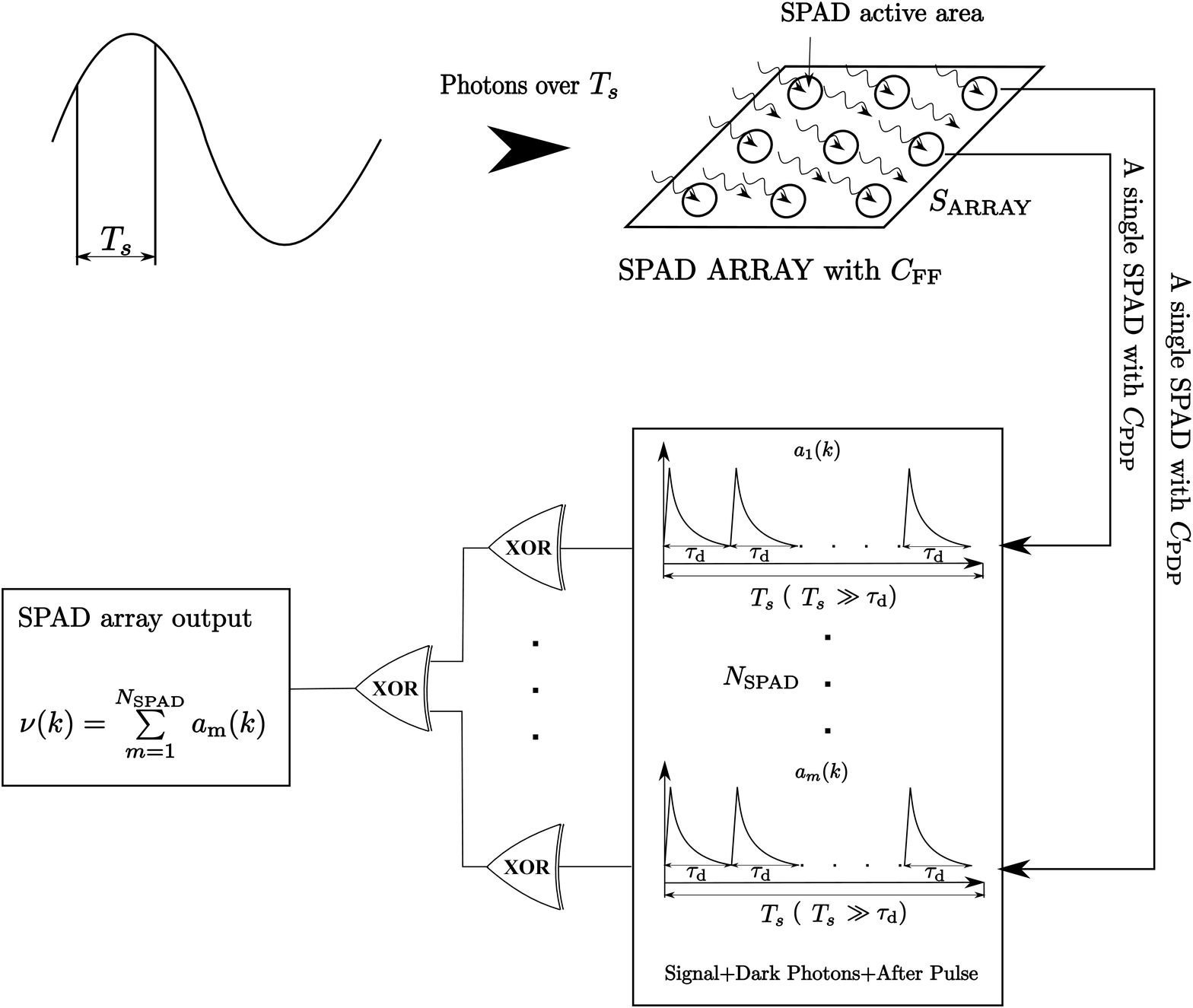}
\caption{Block diagram of photon counting.}
\end{center}
\end{figure}

\subsubsection{Fill Factor (FF)}

FF is the ratio of the total SPAD active area to the total array area. For the SPAD array, FF represents the probability that a photon hits the active area. If the photon triggers an avalanche, it will be counted. In other words, the percentage of photons in a beam light reaching the active area can be approximated to FF. In this study, the value of FF is denoted by $C_{\rm FF}$.

\subsubsection{Photon Detection Probability (PDP)}

PDP is the probability that a photon hitting the active area triggers an avalanche. This avalanche will generate a pulse which can be counted by an accumulator. The accumulator will give the output of the array. PDP is different to the quantum efficiency of the conventional PD, in which the quantum efficiency sometimes includes fill factor effects~\cite{SPADfunda}. In this study, the value of PDP is denoted by $C_{\rm PDP}$.

\subsubsection{Dark Count Rate (DCR)}

A thermally-generated carrier can also trigger an avalanche which increases the array output. Even in complete darkness, this phenomenon still exists as long as the SPAD devices are enabled. The average number of counts in darkness per second is referred to as DCR which is regarded as a fixed signal-unrelated noise of SPAD. In this study, the average DCR of a single SPAD device is denoted by $N_{\rm DCR}$. 

\subsubsection{After Pulsing Probability (APP)}

After pulses are correlated to detections by the time dependent release of trapped carriers~\cite{SPAD2013}. Additional avalanches are triggered after receiving a photon or a dark photon. This means that the after pulsing effect will also increase the array output related to both the incoming signal and the dark counts. The delayed counts will bring inter-symbol interference due to the high data rate. But in low speed transmission, as the sample period is much longer than the delayed time, the after pulsing effect has a negligible effect on the next sample period. In this study, the value of APP is denoted by $P_{\rm AP}$.

\subsubsection{Dead Time}

After an avalanche is triggered, whether caused by the signal photons or dark photons, the SPAD device needs to be actively or passively recharged in a short period of time and this is referred as to dead time. During this time, the SPAD device is unable to detect further signal photons or dark photons. In other words, each individual SPAD in the array can only receive one photon during the dead time. In this study, the value of the dead time is denoted by $\tau_{\rm d}$.

Fig.~2 illustrates the system model of the SPAD array receiving optical signals. In order to generate received O-OFDM symbols, the output of the SPAD array is counted over a symbol duration, $T_{\rm s}$, at time instances $t_k=kT_{\rm s}$ of the received optical signal $\textbf{x}_{\rm{r}}(t)$. These photon counts are denoted by $\nu(k)$ which is the superposition of the photon counts from each individual SPADs, $a_{\rm m}(k)$, as shown in Fig.~2:
\begin{equation}
\nu(k) = \sum\limits_{m = 1}^{N_{\rm SPAD}} a_{\rm m}(k),
\label{sumSPAD}
\end{equation}
where $N_{\rm SPAD}$ is the number of SPAD devices in the array. Generally, as the photon counts from each individual SPAD can be approximately modelled using Poisson statistics, the photon counts at the output of the SPAD array (i.e., $\nu(k)$) can be described by Poisson distribution:
\begin{equation}
{\rm Pr}\Big(\nu(k) = j, \mu(k)\Big) = \exp\Big({-\mu(k)}\Big) \frac{\mu(k)^{j}}{j!},
\label{pois}
\end{equation}
where the average photon counts $\mu(k)$ can be expressed as a function of the received signal and the parameters of the SPADs:
\begin{equation}
\mu(k) = \left[\frac{C_{\rm FF}C_{\rm PDP}}{E_{\rm P}}\int_{t_k}^{t_k+T_{\rm s}}\textbf{x}_{\rm{r}}(t){\rm d}t + n_{\rm DCR}\right](1+P_{\rm AP}),
\label{mu}
\end{equation}
where $E_{\rm P}$ denotes the energy of a photon which is calculated by $\frac{hc_{\rm L}}{w_{\rm L}}$. Note that $h$ denotes Planck's constant; $c_{\rm L}$ is the speed of the light; and $w_{\rm L}$ is the light wavelength of the LED transmitter. The noise caused by dark counts is denoted by $n_{\rm DCR} = N_{\rm DCR}N_{\rm SPAD}T_{\rm s}$. However, when the incoming photon rate is high, Poisson distribution cannot exactly describe the photon counts of SPAD arrays. This is because the dead time effect causes the saturation of SPAD devices and significantly decreases the photon counts. Thus, an exact distribution is also considered and used in this study, and this will be introduced in Section~\uppercase\expandafter{\romannumeral3}.

\subsection{Optical OFDM Demodulation}

The output of the SPAD array is the number of photons ($\nu(k)$), and the system is designed based on a conventional O-OFDM demodulator which requires the amplitude of the electrical signal (optical power) to demodulate the received signal to the original encoded bits. Thus, a photon-to-amplitude equalizer is used to simply convert the received photon number ($\nu(k)$) to the corresponding electrical signal amplitude (optical power), $\textbf{x}'(k)$. The coefficient of the equalizer is calculated by a pilot which can record the effect of the attenuation and the parameters of the SPADs.

Assuming that there is no other distortion effects during the transmission, the recovered signal, $\textbf{x}'(k)$, can be scaled to the original clipped signal, $\textbf{x}_{\rm{clipped}}(k)$. The recovered OFDM symbols from the SPAD are passed through a FFT operation which converts symbols to the frequency domain. In DCO-OFDM, $N/2-1$ symbols are obtained from the corresponding subcarriers to constitute a QAM symbol frame, $\textbf{X}'(n)$. In ACO-OFDM, $N/4$ symbols are obtained. The detected QAM symbols are then decoded by the conventional Maximum Likelihood (ML) estimator in order to obtain the output bit stream.

\section{Nonlinear Distortion in SPAD Receivers}

\begin{figure}[!t]
\begin{center}
\includegraphics[width=0.48\textwidth]{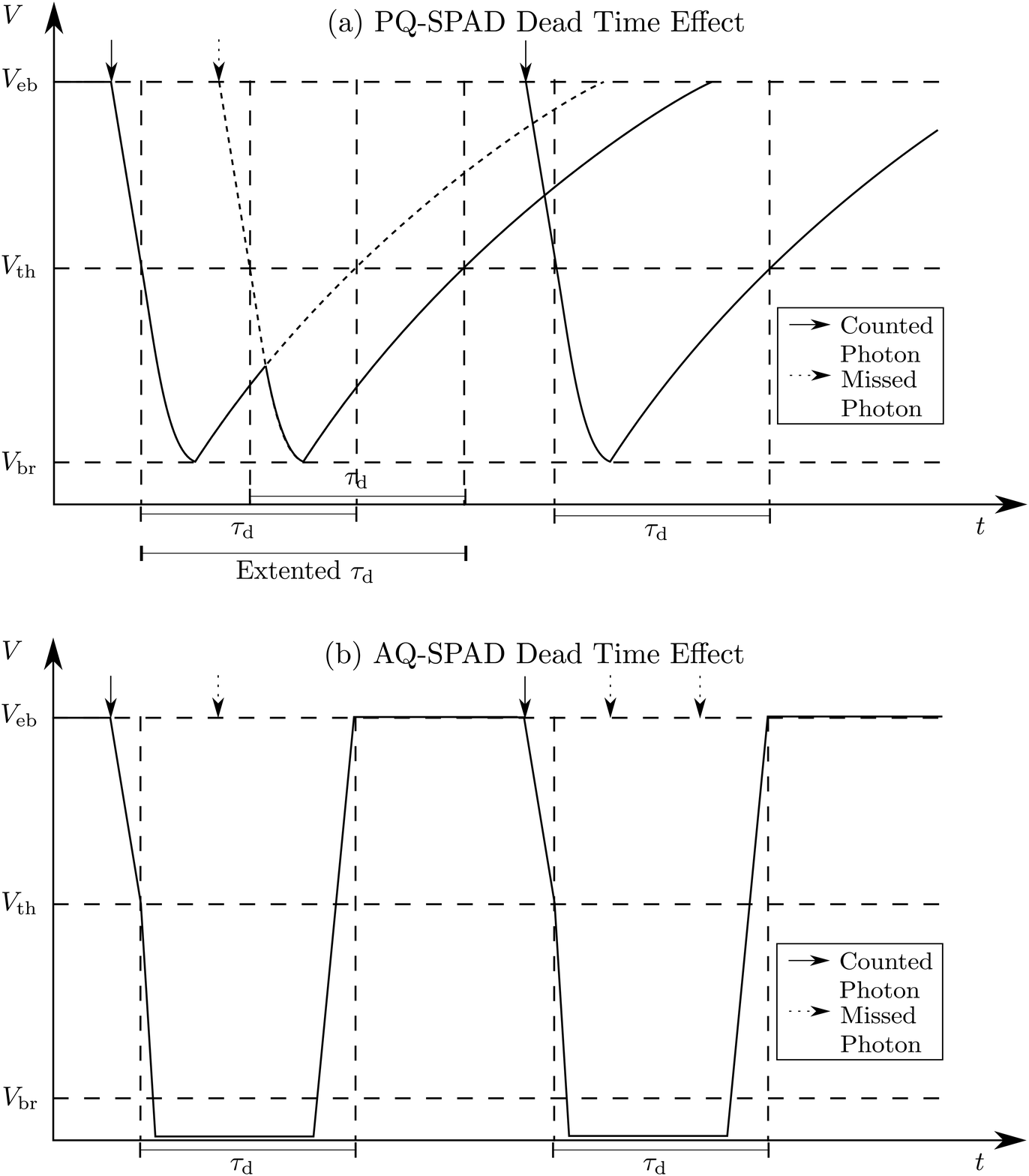}
\caption{(a) PQ SPAD voltage when photons arrive and the dead time is extended by other photons. (b) AQ SPAD voltage when photons arrive and other photons are lost during the dead time.}
\end{center}
\end{figure}

In the SPAD-based system, two SPAD devices with different recharged circuits are applied. The passively recharged SPAD is referred to as passive quenching SPAD (PQ SPAD). The configuration of the passively recharged circuit is presented in~\cite{PQ&AQ}. PQ SPAD is identified as a paralyzable detector where any counts occurring during the dead time (including signal, dark count and after pulse) are not registered but they extend the dead time. As shown in Fig.~3(a), a PQ SPAD device is biased with an excess bias voltage ($V_{\rm eb}$) above the breakdown voltage ($V_{\rm br}$). When an incident photon triggers an avalanche and the SPAD voltage is reduced below $V_{\rm br}$, PQ SPAD is then passively recharged. As soon as the SPAD voltage exceeds $V_{\rm br}$, the PQ SPAD device can be triggered by another photon. As a result, the SPAD voltage remains below the threshold voltage ($V_{\rm th}$). As long as the voltage is lower than $V_{\rm th}$, only the first photon is registered and other photons are lost. This means that the dead time is extended and the PQ-SPAD is paralyzed. Another SPAD device is actively recharged, so-called active quenching SPAD (AQ SPAD). Compared with PQ SPAD, the configuration of AQ SPAD is more complex and requires more area and power~\cite{PQ&AQ}, but when any events arrive during the dead time, the additional events are not registered and do not prolong the dead time. As shown in Fig.~3(b), when an incident photon triggers an avalanche, the voltage of the AQ SPAD device is reduced below $V_{\rm br}$. After the hold-off time (the voltage remaining below $V_{\rm br}$), the SPAD voltage is forcibly returned to $V_{\rm eb}$ by the active recharged circuit. As a consequence, AQ SPAD cannot be triggered by other incident photons during the dead time. As the dead time will not be extended, AQ SPAD is defined as a non-paralyzable detector and has higher count rates than PQ SPAD.

\begin{figure}[!t]
\begin{center}
\includegraphics[width=0.48\textwidth]{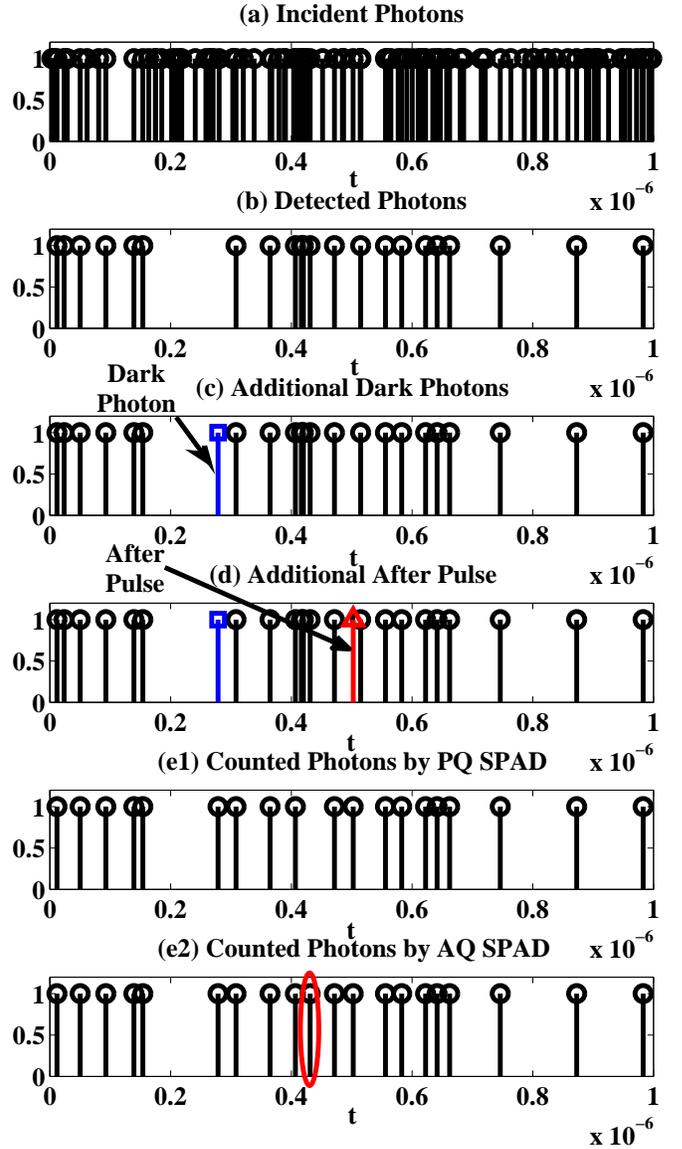}
\caption{Process of photon counting in SPAD devices over $T_{\rm s}$: (a) the original incident photons reaching active area of SPAD devices (related to FF); (b) the detected photons may trigger avalanches (related to PDP); (c) the random dark photons (related to DCR); (d) the after pulses depend on the detected photons and the dark photons (related to APP); (e1) outputs of PQ SPAD; (e2) outputs of AQ SPAD.}
\end{center}
\end{figure}

Fig.~4 shows the process of photon counting in PQ and AQ SPADs over $T_{\rm s}$ which is set to $1 \ {\rm \mu s}$ in the simulation. During $T_{\rm s}$, it is assumed that around 100 photons hit the active area of the SPAD device (Fig.~4(a)). As shown in Fig.~4(b), only a bit of incident photons can trigger avalanches. As noted, the probability of the trigger is PDP. At the same time, the dark current triggers independently Poisson random avalanches (Fig.~4(c)). Afterwards, following the detected photons and the dark photons, the after pulse also provides some additional photon counts (Fig.~4(d)). In PQ SPAD, as only the first triggered avalanche can be recorded during one extended dead time, a limited number of avalanches can be achieved as shown in Fig.~4(e1). Those pulse trains will be registered by the accumulator and the output of a single PQ SPAD device can be obtained. Unlike PQ SPAD, some potential avalanches cannot be triggered during the dead time in AQ SPAD. As the dead time is not extended, the AQ-SPAD device can achieve more photons counts, as shown in Fig.~4(e2). 

\begin{figure}[!t]
\begin{center}
\includegraphics[width=0.48\textwidth]{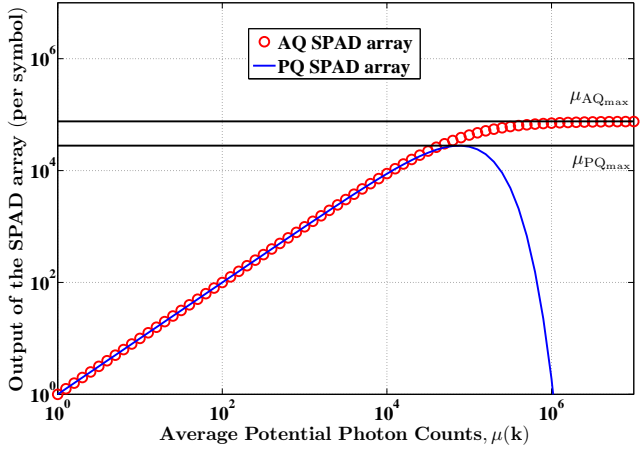}
\caption{Nonlinearity of the PQ SPAD array and the AQ SPAD array.}
\end{center}
\end{figure}

In either PQ SPAD or AQ SPAD, the dead time effect makes a nonlinear reduction on photon counts. For a single PQ SPAD device, the relationship between the average potential counts per second, $\mu_{\rm m}$, and the real photon counts, $\mu_{\rm PQ_m}$, is~\cite{PQ&AQ}:
\begin{equation}
\mu_{\rm PQ_m} = \mu_{\rm m}\exp(-\mu_{\rm m}\tau_{\rm d}).
\end{equation}
Thus, for each $T_{\rm s}$, the average number of the real photon counts, $\mu_{\rm PQ_m}(k)$, is calculated by:
\begin{equation}
\begin{aligned}
\mu_{\rm PQ_m}(k) & = \frac{\mu_{\rm m}(k)}{T_{\rm s}} \exp\left(-\frac{\mu_{\rm m}(k)}{T_{\rm s}}\tau_{\rm d}\right)T_{\rm s} \\
& = \mu_{\rm m}(k) \exp\left(-\frac{\mu_{\rm m}(k)\tau_{\rm d}}{T_{\rm s}}\right).
\end{aligned}
\end{equation}
where $\mu_{\rm m}(k)$ denotes the average potential counts for each single device in the same $T_{\rm s}$. For the SPAD array, $\mu_{\rm m}(k)$ is equal to $\mu(k)/N_{\rm SPAD}$. In this study, if the SPAD array is composed by PQ SPAD devices, the average output of the array during each $T_{\rm s}$ can be expressed as:
\begin{equation}
\begin{aligned}
\mu_{\rm PQ}(k) & = \sum\limits_{m = 1}^{N_{\rm SPAD}} \mu_{\rm PQ_m}(k) \\
& = \mu(k)\exp\left(-\frac{\mu(k)\tau_{\rm d}}{T_{\rm s}N_{\rm SPAD}}\right).
\end{aligned}
\label{PQnon}
\end{equation}
Note that $\mu(k)$ is calculated by (\ref{mu}). According to the process of photon counting, $\mu(k)$ means the average potential counts by the PQ SPAD array. For simplicity, $\mu_{\rm PQ}(k)$ can replace $\mu(k)$ in (\ref{pois}) to estimate the distribution of the SPAD array output. From the nonlinear function of the PQ SPAD array, (\ref{PQnon}), the maximum photon count rate can be calculated:
\begin{equation}
\mu_{\rm PQ_{max}} = \frac{T_{\rm s}N_{\rm SPAD}}{e\tau_{\rm d}},
\end{equation}
where $e$ is Euler's number. As shown in Fig.~5, after reaching $\mu_{\rm PQ_{max}}$, as the PQ SPAD devices are paralyzed, the outputs of the SPAD array rapidly decreases with an increasing rate of incoming photons.

For a single AQ SPAD device, the average real photon counts per second, $\mu_{\rm AQ_m}$, is expressed as a function of $\mu_{\rm m}$~\cite{PQ&AQ}:
\begin{equation}
\mu_{\rm AQ_m} = \frac{\mu_{\rm m}}{1+\mu_{\rm m}\tau_{\rm d}}.
\end{equation}
For each $T_{\rm s}$, the average output of a single device is:
\begin{equation}
\mu_{\rm AQ_m}(k) = \frac{\mu_{\rm m}(k)}{1+\frac{\mu_{\rm m}(k)\tau_{\rm d}}{T_{\rm s}}}.
\end{equation}
Thus the average output of the AQ SPAD array in $T_{\rm s}$ is: 
\begin{equation}
\begin{aligned}
\mu_{\rm AQ}(k) & = \sum\limits_{m = 1}^{N_{\rm SPAD}} \mu_{\rm AQ_m}(k) \\
& = \frac{\mu(k)}{1+\frac{\mu(k)\tau_{\rm d}}{T_{\rm s}N_{\rm SPAD}}}.
\end{aligned}
\label{AQnon}
\end{equation}
Thus, the maximum photon count rate of the AQ SPAD array is:
\begin{equation}
\mu_{\rm AQ_{max}} = \frac{T_{\rm s}N_{\rm SPAD}}{\tau_{\rm d}}.
\end{equation}
As shown in Fig.~5, when the incoming photon rate increases, the AQ SPAD devices are non-paralyzed but the outputs dramatically converge to $\mu_{\rm AQ_{max}}$. In other words, if the average potential photon counts, including signal photons, dark photons and after pulse counts, are more than $\mu_{\rm AQ_{max}}$, the AQ SPAD array will be saturated. The photon counts at the output of the SPAD array are constrained to $\mu_{\rm AQ_{max}}$ and the extra photons are refused and lost.

\begin{figure}[!t]
\begin{center}
\includegraphics[width=0.48\textwidth]{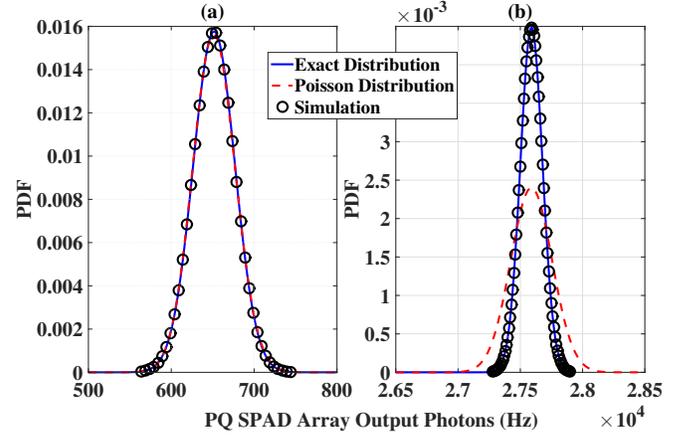}
\caption{The probability density functions of the PQ SPAD array output. The exact distribution, Poisson distribution and simulation results are compared over $T_{\rm s} = 1 \mu s$: (a) the number of the total incident photons is $10^4$; (b) the number of the total incident photons is $10^6$.}
\end{center}
\end{figure}

As mentioned in Section~\uppercase\expandafter{\romannumeral2}, the photon count distribution of SPAD receivers can be approximately described by Poisson distribution. For PQ SPAD, as shown in Fig.~6(a), when the total incident photons are $10^4$ over $T_{\rm s}$, the Possion distribution can accurately describe the real simulated distribution. However, when incident photons increase to $10^6$ (Fig.~6(b)), the variance of the Poisson distribution is too high to describe the distribution of output photons. Thus, in order to get more accurate results, an exact distribution is used to replace the Poisson distribution~\cite{SPADreceiver}:
\begin{equation}
\begin{aligned}
{\rm Pr}_{\rm PQ}(a,\mu_{\rm m}) = \sum_{j=a}^{a_{\rm PQ_{max}}-1} & {j \choose a}(-1)^{(j-a)}\frac{\mu_{\rm m}^j}{j!} \\
& \exp(-j\mu_{\rm m}\tau_{\rm d})(T_{\rm s}-j\tau_{\rm d})^j.
\end{aligned}
\end{equation}
Note that ${\rm Pr}_{\rm PQ}(a,\mu_{\rm m})$ is the exact photon count distribution of a single PQ SPAD during $T_{\rm s}$. The maximum photon count rate for a single PQ SPAD device is denoted by $a_{\rm PQ_{max}}$ which is equal to $\lfloor T_{\rm s}/e\tau_{\rm d} \rfloor$. In the PQ SPAD array, it is assumed that the photon count distributions of each single device are the same during $T_{\rm s}$. The distribution can be written as a vector:
\begin{equation}
\begin{aligned}
\textbf{Pr}_{\rm m}(k) = & \Bigg[{\rm Pr}_{\rm PQ}\left(0,\frac{\mu(k)}{T_{\rm s}N_{\rm SPAD}}\right), {\rm Pr}_{\rm PQ}\left(1,\frac{\mu(k)}{T_{\rm s}N_{\rm SPAD}}\right) , ..., \\ 
 & {\rm Pr}_{\rm PQ}\left(a_{\rm PQ_{max}}-1,\frac{\mu(k)}{T_{\rm s}N_{\rm SPAD}}\right)\Bigg].
\end{aligned}
\label{Prarray}
\end{equation}
Thus, according to (\ref{sumSPAD}), the joint distribution of the whole SPAD array can be calculated as:
\begin{equation}
\textbf{Pr}(k) = \underbrace{\textbf{Pr}_{\rm m}(k)*\textbf{Pr}_{\rm m}(k)*...*\textbf{Pr}_{\rm m}(k)}_{(N_{\rm SPAD}-1) \ \rm{times \  convolution}}.
\label{arraydis}
\end{equation}
According to~\cite{SPADreceiver} and (\ref{arraydis}), the exact expectation of the PQ SPAD array output during $T_{\rm s}$ is:
\begin{equation}
\begin{aligned}
E_{\rm PQ}(k) & = N_{\rm SPAD}\mu_{\rm m}\exp(-\mu_{\rm m}\tau_{\rm d})(T_{\rm s}-\tau_{\rm d}) \\
& \approx N_{\rm SPAD}\frac{\mu(k)}{T_{\rm s}N_{\rm SPAD}} \exp(-\frac{\mu(k)}{T_{\rm s}N_{\rm SPAD}}\tau_{\rm d}) T_{\rm s} \\
& = \mu(k)\exp\left(-\frac{\mu(k)\tau_{\rm d}}{T_{\rm s}N_{\rm SPAD}}\right) =  \mu_{\rm PQ}(k).\\
\end{aligned}
\end{equation}
Note that the symbol period, $T_{\rm s}$, is assumed to be much longer than the dead time, $\tau_{\rm d}$, in this study. Therefore, the exact expectation can be approximated to the average output of the array in (\ref{PQnon}). The exact variance of the array output is:
\begin{equation}
\begin{aligned}
\sigma^2_{\rm PQ}(k) = & N_{\rm SPAD}\bigg[ \mu_{\rm m}^2\exp(-2\mu_{\rm m}\tau_{\rm d})(3\tau_{\rm d}^2-2T_{\rm s}\tau_{\rm d}) \\ & +\mu_{\rm m}\exp(-\mu_{\rm m}\tau_{\rm d})T_{\rm s} \bigg].
\label{vPQe}
\end{aligned}
\end{equation}
As shown in Fig.~6(a) and (b),  the exact distribution is well matched with the simulation distribution. As a result, the exact distribution performs better on describing the photon counts than the Poisson distribution in the PQ SPAD array.

\begin{figure}[!t]
\begin{center}
\includegraphics[width=0.48\textwidth]{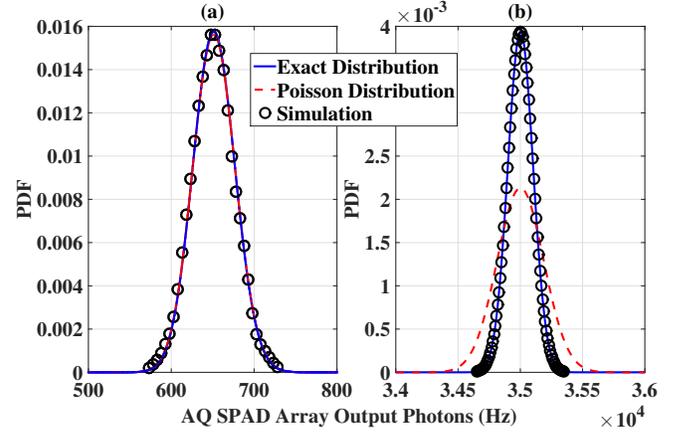}
\caption{The probability density functions of the AQ SPAD array output. The exact distribution, Poisson distribution and simulation results are compared over $T_{\rm s} = 1 \mu s$: (a) the number of the total incident photons is $10^4$; (b) the number of the total incident photons is $10^6$.}
\end{center}
\end{figure}

\begin{figure*}[tp]
\setcounter{Mycount}{\value{equation}} % Store the current equation number.
\setcounter{equation}{31}
\begin{equation}
\begin{aligned}
{\rm E}&\left[N(x)z_{\rm PQ}\Big(N(x)\Big)\right] = \int_{-\infty}^{\infty}N(x)z_{\rm PQ}\Big(N(x)\Big)\frac{1}{\sigma_{\rm x}}\phi\left(\frac{x-\rho}{\sigma_{\rm x}}\right){\rm d}x %+ \int_{-\infty}^{0}N(0)z_{\rm PQ}\Big(N(0)\Big)\frac{1}{\sigma_{\rm x}}\phi\left(\frac{x-\rho}{\sigma_{\rm x}}\right){\rm d}x 
\\
& = \int_{0}^{\infty}(C_{\rm s}x+C_{\rm n})^2 \exp\left(\frac{1}{2}C_{\rm t}^2C_{\rm s}^2\sigma_{\rm x}^2-C_{\rm t}C_{\rm s}\rho-C_{\rm t}C_{\rm n}\right) \frac{1}{\sigma_{\rm x}}\phi\left(\frac{x-\rho+C_{\rm t}C_{\rm s}\sigma_{\rm x}^2}{\sigma_{\rm x}}\right){\rm d}x + C_{\rm n}^2\exp(-C_{\rm t}C_{\rm n})Q\left(\frac{\rho}{\sigma_{\rm x}}\right)\\
& = \exp\left(\frac{1}{2}C_{\rm t}^2C_{\rm s}^2\sigma_{\rm x}^2-C_{\rm t}C_{\rm s}\rho-C_{\rm t}C_{\rm n}\right)
\Bigg \{  
C_{\rm s}^2\bigg[\left(\rho-C_{\rm t}C_{\rm s}\sigma_{\rm x}^2\right)^2Q\left(\frac{C_{\rm t}C_{\rm s}\sigma_{\rm x}^2-\rho}{\sigma_{\rm x}}\right)  +  \sigma_{\rm x}^2Q\left(\frac{C_{\rm t}C_{\rm s}\sigma_{\rm x}^2-\rho}{\sigma_{\rm x}}\right) \\ & \quad +  \left(\rho-C_{\rm t}C_{\rm s}\sigma_{\rm x}^2\right)\sigma_{\rm x}\phi\left(\frac{C_{\rm t}C_{\rm s}\sigma_{\rm x}^2-\rho}{\sigma_{\rm x}}\right)\bigg]
+ 2C_{\rm s}C_{\rm n}\bigg[\left(\rho-C_{\rm t}C_{\rm s}\sigma_{\rm x}^2\right)Q\left(\frac{C_{\rm t}C_{\rm s}\sigma_{\rm x}^2-\rho}{\sigma_{\rm x}}\right)  +  \sigma_{\rm x}\phi\left(\frac{C_{\rm t}C_{\rm s}\sigma_{\rm x}^2-\rho}{\sigma_{\rm x}}\right)\bigg] \\
& \quad+ C_{\rm n}^2Q\left(\frac{\rho-C_{\rm t}C_{\rm s}\sigma_{\rm x}^2}{\sigma_{\rm x}}\right) \Bigg\} 
+ C_{\rm n}^2\exp(-C_{\rm t}C_{\rm n})Q\left(\frac{\rho}{\sigma_{\rm x}}\right).
\end{aligned}
\label{ENzPQ}
\end{equation}
\begin{equation}
\begin{aligned}
{\rm E}&\left[N^2(x)\right] = \int_{-\infty}^{\infty}N^2(x)\frac{1}{\sigma_{\rm x}}\phi\left(\frac{x-\rho}{\sigma_{\rm x}}\right){\rm d}x %+ \int_{-\infty}^{0}N^2(0)\frac{1}{\sigma_{\rm x}}\phi\left(\frac{x-\rho}{\sigma_{\rm x}}\right){\rm d}x 
\\
& = C_{\rm s}^2\left[ \rho^2Q\left(-\frac{\rho}{\sigma_{\rm x}}\right) + \sigma_{\rm x}^2Q\left(-\frac{\rho}{\sigma_{\rm x}}\right) + \rho\sigma_{\rm x}\phi\left(\frac{\rho}{\sigma_{\rm x}}\right) \right] + 2C_{\rm s}C_{\rm n}\left[ \rho Q\left(-\frac{\rho}{\sigma_{\rm x}}\right) + \sigma_{\rm x}\phi\left(\frac{\rho}{\sigma_{\rm x}}\right) \right] + C_{\rm n}^2.
\end{aligned}
\label{EN2PQ}
\end{equation}
\setcounter{Mycount2}{\value{equation}}
\setcounter{equation}{\value{Mycount}}
\hrulefill
\end{figure*}

For AQ SPAD, in order to achieve more accurate results, an exact distribution is also used to replace the Poisson distribution~\cite{DeadTime}:
\noindent (1) for $0 \leqslant a \leqslant a_{\rm AQ_{max}}-1$ :
\begin{equation}
\begin{aligned}
{\rm Pr}_{\rm AQ}(a,\mu_{\rm m}) = & \lambda \Bigg[ \sum_{j = 0}^{a-2}(a-1-j){\rm Pr}(j, S_{a-1}) \\ 
& - 2 \sum_{j = 0}^{a-1}(a-j){\rm Pr}(j, S_{a}) \\
& + \sum_{j = 0}^{a}(a+1-j){\rm Pr}(j, S_{a+1}) \Bigg],
\end{aligned}
\end{equation}
\noindent (2) for $a = a_{\rm AQ_{max}}$ :
\begin{equation}
\begin{aligned}
{\rm Pr}_{\rm AQ}(a,\mu_{\rm m}) = & \lambda \Bigg[ \sum_{j = 0}^{a-2}(a-1-j){\rm Pr}(j, S_{a-1}) \\ 
& - 2\sum_{j = 0}^{a-1}(a-j){\rm Pr}(j, S_{a}) - \mu_{\rm m}T_{\rm s}\Bigg]+a+1,
\end{aligned}
\end{equation}
\noindent (3) for $a = a_{\rm AQ_{max}}+1$ :
\begin{equation}
\begin{aligned}
{\rm Pr}_{\rm AQ}(a,\mu_{\rm m}) = & \lambda \Bigg[ \sum_{j = 0}^{a-2}(a-1-j){\rm Pr}(j, S_{a-1}) \\
& + \mu_{\rm m}T_{\rm s}\Bigg] - a+1.
\end{aligned}
\end{equation}
Note that ${\rm Pr}_{\rm AQ}(a,\mu_{\rm m})$ is the exact photon count distribution of a single AQ SPAD during $T_{\rm s}$. The maximum photon count rate for a single AQ SPAD device is denoted by $a_{\rm AQ_{max}}$ which is equal to $\lfloor T_{\rm s}/\tau_{\rm d} \rfloor$; $\lambda = (1+\mu_{\rm m}\tau_{\rm d})^{-1}$; and $S_a = \mu_{\rm m}(T_{\rm s}-a\tau_{\rm d})$. According to (\ref{Prarray}), (\ref{arraydis}) and~\cite{DeadTime}, when $T_{\rm s}$ is much longer than $\tau_{\rm d}$, the exact expectation of the AQ SPAD array output is:
\begin{equation}
\begin{aligned}
E_{\rm AQ}(k) & = N_{\rm SPAD}E_{\rm m}(k) = N_{\rm SPAD}\lambda\mu_{\rm m}T_{\rm s} \\
& = N_{\rm SPAD}\left[1+\frac{\mu(k)}{T_{\rm s}N_{\rm SPAD}}\tau_{\rm d}\right]^{-1}\frac{\mu(k)}{T_{\rm s}N_{\rm SPAD}}T_{\rm s} \\
& = \frac{\mu(k)}{1+\frac{\mu(k)\tau_{\rm d}}{T_{\rm s}N_{\rm SPAD}}} =  \mu_{\rm AQ}(k),\\
\end{aligned}
\end{equation}
It can be seen that the exact photon count distribution of the AQ SPAD array has the same mean value as the Poisson distribution from (\ref{AQnon}). According to the variance calculation of a single AQ SPAD device~\cite{DeadTime}, the exact variance of the AQ array output is:
\begin{equation}
\begin{aligned}
\sigma^2_{\rm AQ}(k) & = N_{\rm SPAD}\sigma^2_{\rm m}(k) \\
& = N_{\rm SPAD}\lambda^3\left[\mu_{\rm m}T_{\rm s} + g^2\lambda(1+\frac{2}{3}g+\frac{1}{6}g^2)\right],
\label{vAQe}
\end{aligned}
\end{equation}
where $g = \mu_{\rm m}\tau_{\rm d}$. Fig.~7 shows the photon count distribution of an AQ SPAD array. When the number of the total incident photons is low ($10^4$), the simulation result closely matches with both Poisson distribution and the exact distribution. Furthermore, as a result of a comparison between Fig.~6(a) and Fig.~7(a), it shows that the PQ and AQ SPAD array have the similar photon count distribution. This is because the linear region of the PQ SPAD array is almost coincident with the AQ SPAD array when the photon rate is low (Fig.~5). In Fig.~7(b), compared with Poisson distribution, the exact distribution is closer to the simulation result. Moreover, when compared with the photon count distribution of the PQ SPAD array (Fig.~6(b)), the AQ SPAD array has a higher mean value of photon counts when the nonlinear distortion occurs.

\section{Theoretical Analysis of SPAD-based OFDM}

The analytical BER performance of SPAD-based OFDM is presented in this section. In the SPAD-based OFDM system, some high amplitude symbols in the recovered signal ($\textbf{x}'(k)$) are distorted by PQ and AQ recharged circuits resulting in loss of information. This causes a unique receiver nonlinear distortion which should be considered in the theoretical analysis. According to~\cite{dtv0001,dsh2013non}, a nonlinear distortion in an OFDM based system can be described with a gain factor ($\alpha$) and additional noise ($Y$), both of which can be explained and quantified with the Bussgang theorem. It states that if an independent Gaussian random variable, $X$, passes through a nonlinear transformation, $z(X)$, then~\cite{tech:jbuss01}:
\begin{equation}
\left\{
\begin{array}{lll}
z(X) = \alpha X + Y, \\
{\rm E}[XY] = 0,
\end{array}
\right.
\label{BT}
\end{equation}
where $\alpha$ is a constant which can be derived as:
\begin{equation}
\alpha = \frac{{\rm E}\left[Xz(X)\right]}{{\rm E}\left[X^2\right]}.
\label{alpha}
\end{equation}
According to (\ref{BT}), the variance of the additional noise, $\sigma_{\rm Y}^2$, can be calculated by:
\begin{equation}
\sigma_{\rm Y}^2 = {\rm E}\left[Y^2\right] - {\rm E}^2[Y],
\label{sy}
\end{equation}
where:
\begin{equation}
{\rm E}\left[Y^2\right] = {\rm E}\left[z^2(x)\right] - {\rm E}\left[\alpha^2X^2\right],
\label{ey2}
\end{equation}
\begin{equation}
{\rm E}[Y] = {\rm E}[z(x)] - {\rm E}[\alpha X].
\label{ey}
\end{equation}
To describe the analytical BER calculations of SPAD-based OFDM, the following formulas are defined. The standard normal distribution probability density function (PDF) is: 
\begin{equation}
\phi(x) = \frac{1}{\sqrt{2\pi}} \exp\left(-\frac{x^2}{2}\right).
\end{equation}
According to (\ref{mu}), the relationship between each Gaussian random variable, $x$, and the related number of photons, $N(x)$, is:
\begin{equation}
N(x) = \left\{
\begin{array}{ll}
C_{\rm s}x + C_{\rm n}, & x \geq 0, \\
0, & x<0,
\end{array}
\right.
\end{equation}
where $C_{\rm s} = C_{\rm FF}C_{\rm PDP}PT_{\rm s}(1+P_{\rm AP})/E_{\rm P}$ and $C_{\rm n} = n_{\rm DCR}(1+P_{\rm AP})$. Note that $P$ is the average received optical power.
 
\subsection{OFDM with PQ SPAD}

\begin{figure*}
\setcounter{Mycount}{\value{equation}} % Store the current equation number.
\setcounter{equation}{33}
\begin{equation}
\begin{aligned}
{\rm E}&\left[z_{\rm PQ}^2\Big(N(x)\Big)\right] = \int_{-\infty}^{\infty}z_{\rm PQ}^2\Big(N(x)\Big)\frac{1}{\sigma_{\rm x}}\phi\left(\frac{x-\rho}{\sigma_{\rm x}}\right){\rm d}x %+ \int_{-\infty}^{0}z_{\rm PQ}^2\Big(N(0)\Big)\frac{1}{\sigma_{\rm x}}\phi\left(\frac{x-\rho}{\sigma_{\rm x}}\right){\rm d}x 
\\
& = \int_{0}^{\infty}(C_{\rm s}x+C_{\rm n})^2 \exp\left(2C_{\rm t}^2C_{\rm s}^2\sigma_{\rm x}^2-2C_{\rm t}C_{\rm s}\rho - 2C_{\rm t}C_{\rm n}\right) \frac{1}{\sigma_{\rm x}}\phi\left(\frac{x-\rho+2C_{\rm t}C_{\rm s}\sigma_{\rm x}^2}{\sigma_{\rm x}}\right){\rm d}x + C_{\rm n}^2\exp(-2C_{\rm t}C_{\rm n})Q\left(\frac{\rho}{\sigma_{\rm x}}\right) \\
& = \exp\left(2C_{\rm t}^2C_{\rm s}^2\sigma_{\rm x}^2-2C_{\rm t}C_{\rm s}\rho-2C_{\rm t}C_{\rm n}\right)
\Bigg \{  
C_{\rm s}^2\bigg[\left(\rho-2C_{\rm t}C_{\rm s}\sigma_{\rm x}^2\right)^2Q\left(\frac{2C_{\rm t}C_{\rm s}\sigma_{\rm x}^2-\rho}{\sigma_{\rm x}}\right)  +  \sigma_{\rm x}^2Q\left(\frac{2C_{\rm t}C_{\rm s}\sigma_{\rm x}^2-\rho}{\sigma_{\rm x}}\right) \\ & \quad +  \left(\rho-2C_{\rm t}C_{\rm s}\sigma_{\rm x}^2\right)\sigma_{\rm x}\phi\left(\frac{2C_{\rm t}C_{\rm s}\sigma_{\rm x}^2-\rho}{\sigma_{\rm x}}\right)\bigg]
+ 2C_{\rm s}C_{\rm n}\bigg[\left(\rho-2C_{\rm t}C_{\rm s}\sigma_{\rm x}^2\right)Q\left(\frac{2C_{\rm t}C_{\rm s}\sigma_{\rm x}^2-\rho}{\sigma_{\rm x}}\right)  +  \sigma_{\rm x}\phi\left(\frac{2C_{\rm t}C_{\rm s}\sigma_{\rm x}^2-\rho}{\sigma_{\rm x}}\right)\bigg] \\
& \quad+ C_{\rm n}^2Q\left(\frac{\rho-2C_{\rm t}C_{\rm s}\sigma_{\rm x}^2}{\sigma_{\rm x}}\right) \Bigg\} 
+ C_{\rm n}^2\exp(-2C_{\rm t}C_{\rm n})Q\left(\frac{\rho}{\sigma_{\rm x}}\right).
\end{aligned}
\label{Ez2PQ}
\end{equation}
\begin{equation}
\begin{aligned}
{\rm E}&\left[z_{\rm PQ}\Big(N(x)\Big)\right] = \int_{-\infty}^{\infty}z_{\rm PQ}\Big(N(x)\Big)\frac{1}{\sigma_{\rm x}}\phi\left(\frac{x-\rho}{\sigma_{\rm x}}\right){\rm d}x %+ \int_{-\infty}^{0}z_{\rm PQ}\Big(N(0)\Big)\frac{1}{\sigma_{\rm x}}\phi\left(\frac{x-\rho}{\sigma_{\rm x}}\right){\rm d}x 
\\
& = \int_{0}^{\infty}(C_{\rm s}x+C_{\rm n})\exp\left(\frac{1}{2}C_{\rm t}^2C_{\rm s}^2\sigma_{\rm x}^2-C_{\rm t}C_{\rm s}\rho-C_{\rm t}C_{\rm n}\right) \frac{1}{\sigma_{\rm x}}\phi\left(\frac{x-\rho+C_{\rm t}C_{\rm s}\sigma_{\rm x}^2}{\sigma_{\rm x}}\right){\rm d}x + C_{\rm n}\exp(-C_{\rm t}C_{\rm n})Q\left(\frac{\rho}{\sigma_{\rm x}}\right)\\
& = \exp\left(\frac{1}{2}C_{\rm t}^2C_{\rm s}^2\sigma_{\rm x}^2-C_{\rm t}C_{\rm s}\rho-C_{\rm t}C_{\rm n}\right)
\Bigg \{ 
C_{\rm s}\bigg[\left(\rho-C_{\rm t}C_{\rm s}\sigma_{\rm x}^2\right)Q\left(\frac{C_{\rm t}C_{\rm s}\sigma_{\rm x}^2-\rho}{\sigma_{\rm x}}\right)  +  \sigma_{\rm x}\phi\left(\frac{C_{\rm t}C_{\rm s}\sigma_{\rm x}^2-\rho}{\sigma_{\rm x}}\right)\bigg] \\
& \quad+ C_{\rm n}Q\left(\frac{\rho-C_{\rm t}C_{\rm s}\sigma_{\rm x}^2}{\sigma_{\rm x}}\right) \Bigg\}
+ C_{\rm n}\exp(-C_{\rm t}C_{\rm n})Q\left(\frac{\rho}{\sigma_{\rm x}}\right).
\end{aligned}
\label{EzPQ}
\end{equation}
\begin{equation}
{\rm E}\left[N(x)\right] = \int_{-\infty}^{\infty}N(x)\frac{1}{\sigma_{\rm x}}\phi\left(\frac{x-\rho}{\sigma_{\rm x}}\right){\rm d}x = C_{\rm s}\bigg[\rho Q\left(-\frac{\rho}{\sigma_{\rm x}}\right)+\sigma_{\rm x}\phi\left(\frac{\rho}{\sigma_{\rm x}}\right)\bigg]+C_{\rm n}. %+ \int_{-\infty}^{0}N(0)\frac{1}{\sigma_{\rm x}}\phi\left(\frac{x-\rho}{\sigma_{\rm x}}\right){\rm d}x
\label{ENPQ}
\end{equation}
\setcounter{Mycount3}{\value{equation}}
\setcounter{equation}{\value{Mycount}}
\hrulefill
\end{figure*}

According to (\ref{PQnon}) and (\ref{BT}), the nonlinear transformation function of PQ SPAD OFDM is:
\begin{equation}
z_{\rm PQ}\Big(N(x)\Big) = N(x)\exp(-C_{\rm t}N(x)) = \alpha_{\rm PQ}N(x) +Y_{\rm PQ},
\end{equation}
where $C_{\rm t} = \tau_{\rm d} / (T_{\rm s}N_{\rm SPAD})$. Thus, to calculate the nonlinear gain factor of PQ SPAD OFDM, the components in (\ref{alpha}), ${\rm E}\left[N(x)z_{\rm PQ}\big(N(x)\big)\right]$ and ${\rm E}\left[N^2(x)\right]$, are derived by (\ref{ENzPQ}) and (\ref{EN2PQ}). Note that $\sigma_{\rm x}$ and $\rho$ respectively denote the standard deviation and mean value of the bipolar normalized OFDM symbols. As a result, a closed-form expression of $\alpha_{\rm PQ}$ can be obtained. Moreover, ${\rm E}\left[z_{\rm PQ}^2\big(N(x)\big)\right]$, ${\rm E}\left[z_{\rm PQ}\big(N(x)\big)\right]$ and ${\rm E}\left[N(x)\right]$ are calculated by (\ref{Ez2PQ}), (\ref{EzPQ}) and (\ref{ENPQ}). Thus, the variance of the additional distortion noise in (\ref{sy}) can be derived as a closed-form expression. 

In this study, the nonlinear distortion effects in PQ SPAD-based ACO-OFDM and DCO-OFDM are discussed separately as follows.

\subsubsection{PQ SPAD ACO-OFDM} 
In ACO-OFDM, the standard deviation of the original bipolar OFDM symbols, ${\textbf x}(k)$, is:
\setcounter{equation}{\value{Mycount3}}
\begin{equation}
\sigma_{\rm m-ACO} = \sqrt{\frac{M-1}{3}},
\end{equation}
where $M$ is the constellation size of QAM symbols. In this study, DC bias is not assumed in ACO-OFDM. As the negative part of the OFDM frame is clipped, the remaining part can be described by a half Gaussian distribution with zero mean. Thus, the mean value of the remaining part is:
\begin{equation}
{\rm E}[\textbf{x}_{\rm{clipped}}(k)] = \frac{\sigma_{\rm m-ACO}}{\sqrt{2\pi}}.
\end{equation}
As noted, the mean value of the transmitted symbols is normalized to one in the simulation. As a result, the intervals of the normalized ACO-OFDM symbols can be specified from 0 to $\infty$ in a zero-mean Gaussian distribution ($\rho = 0$) with a standard deviation:
\begin{equation}
\sigma_{\rm x-ACO} = \frac{\sigma_{\rm m-ACO}}{{\rm E}[\textbf{x}_{\rm{clipped}}(k)]} = \sqrt{2\pi}.
\end{equation}
Then, based on (\ref{alpha}), (\ref{ENzPQ}) and (\ref{EN2PQ}), the nonlinear gain factor of PQ SPAD ACO-OFDM can be derived as:
\begin{equation}
\alpha_{\rm PQ-ACO} = \left. \frac{{\rm E}\left[N(x)z_{\rm PQ}\Big(N(x)\Big)\right]}{{\rm E}\left[N(x)^2\right]} \right |_{\rho = 0, \sigma_{\rm x} =  \sqrt{2\pi}}. \\
\label{aPQACO}
\end{equation}
According to (\ref{EN2PQ}-\ref{ENPQ}) and (\ref{aPQACO}), the equations, (\ref{ey2}) and (\ref{ey}), become:
\begin{equation}
{\rm E}\left[Y^2\right] = {\rm E}\left[z_{\rm PQ}\Big(N(x)\Big)^2\right] - \alpha_{\rm PQ-ACO}^2{\rm E}\left[N(x)^2\right],
\end{equation}
\begin{equation}
{\rm E}[Y] = {\rm E}\left[z_{\rm PQ}\Big(N(x)\Big)\right] - \alpha_{\rm PQ-ACO}{\rm E}\left[N(x)\right],
\end{equation}
where $\rho = 0$ and $\sigma_{\rm x} =  \sqrt{2\pi}$. Thus, the variance of the additional noise in PQ SPAD ACO-OFDM, $\sigma_{\rm Y-PQ-ACO}^2$, can be obtained by (\ref{sy}). Since the number of OFDM subcarriers and SPAD devices is high enough, the resulting variances can be approximated to variances of Gaussian distribution. Thus, according to~\cite{dsh2013non}, the resulting SNR through the nonlinear transformation can be calculated with the following formula:
\begin{equation}
{\rm SNR}^{\rm PQ}_{\rm ACO} = \frac{\alpha_{\rm PQ-ACO}^2C_{\rm s}^2\sigma_{\rm x-ACO}^2}{2R_{\rm ACO}(\sigma_{\rm Y-PQ-ACO}^2 + \sigma_{\rm N-PQ}^2)}.
\end{equation}
Note that $R_{\rm ACO}$ is the spectral efficiency of ACO-OFDM which is $\frac{1}{4}\log_2(M)$ and $\sigma_{\rm N-PQ}^2$ is the variance of the shot noise which is related to the received signals. In the case of \textit{Poisson distribution}, the variance is equal to the mean value. Thus, for each received symbols, $\sigma_{\rm N-PQ}^2(x)$ is equal to $z_{\rm PQ}\left(N(x)\right)$. As a result, $\sigma_{\rm N-PQ}^2$ can be derived as:
\begin{equation}
\sigma_{\rm N-PQ}^2 = {\rm E}\left[\sigma_{\rm N-PQ}^2(x)\right] = {\rm E}\left[z_{\rm PQ}\Big(N(x)\Big)\right].
\label{sNPQ}
\end{equation}
According to ($\ref{EzPQ}$), the value of $\sigma_{\rm N-PQ}^2$ can be obtained. Note that $\sigma_{\rm x}$ is set to $\sqrt{2\pi}$ and $\rho$ is set to 0 in PQ SPAD ACO-OFDM. In the case of the \textit{exact distribution}, the variance of the shot noise component in the PQ SPAD array can be calculated according to (\ref{vPQe}):
\begin{equation}
\begin{aligned}
& \sigma^2_{\rm N-PQ} = {\rm E}\left[\sigma_{\rm PQ}^2(x)\right] \\
& = {\rm E}\left[ \frac{3\tau_{\rm d}-2T_{\rm s}}{T_{\rm s}}C_{\rm t}z^2_{\rm PQ}\Big(N(x)\Big) + z_{\rm PQ}\Big(N(x)\Big)  \right].
\end{aligned}
\end{equation}
Finally, based on the conventional BER calculation of O-OFDM~\cite{dsh1102}, the analytical BER performance of PQ SPAD ACO-OFDM can be derived by:
\begin{equation}
\begin{aligned}
{\rm BER_{PQ-ACO}} = \frac{4(\sqrt{M}-1)}{\sqrt{M}\log_2(M)}Q\left(\sqrt{\frac{3R_{\rm ACO}}{M-1}{\rm SNR^{\rm PQ}_{\rm ACO}}}\right) \\ +\frac{4(\sqrt{M}-2)}{\sqrt{M}\log_2(M)}Q\left(3\sqrt{\frac{3R_{\rm ACO}}{M-1}{\rm SNR^{\rm PQ}_{\rm ACO}}}\right).
\end{aligned}
\label{BERPA}
\end{equation}

\subsubsection{PQ SPAD DCO-OFDM}

In DCO-OFDM, the standard deviation of the original bipolar OFDM symbols, ${\textbf x}(k)$, is:
\begin{equation}
\sigma_{\rm m-DCO} = \sqrt{\frac{2(M-1)(N-2)}{3N}}.
\end{equation}
As a DC bias component is added to the original bipolar OFDM symbols and some information-carrying symbols are clipped, a clipping distortion noise has to be considered in DCO-OFDM. Based on (\ref{Bdc}), the DC bias in DCO-OFDM is:
\begin{equation}
B_{\rm DC} = \beta\sigma_{\rm m-DCO}
\end{equation}
According to (\ref{alpha}), the clipping distortion factor of DCO-OFDM can be derived as:
\begin{equation}
\begin{aligned}
\alpha_{\rm c} & = \frac{\int_{0}^{\infty} \frac{x^2}{\sigma_{\rm m-DCO}}\phi\left(\frac{x-B_{\rm DC}}{\sigma_{\rm m-DCO}}\right){\rm d}x}{\sigma_{\rm m-DCO}^2+B_{\rm DC}^2} \\
& = Q\left(-\frac{B_{\rm DC}}{\sigma_{\rm m-DCO}}\right) + \beta G_{\rm DC} \phi\left(\frac{B_{\rm DC}}{\sigma_{\rm m-DCO}}\right),
\end{aligned}
\end{equation}
where $G_{\rm DC}$ denotes the attenuation of the original signal power, $\sigma_{\rm m-DCO}^2$, due to the DC bias in DCO-OFDM. It is defined as:
\begin{equation}
G_{\rm DC} = \frac{\sigma_{\rm m-DCO}^2}{\sigma_{\rm m-DCO}^2 + B_{\rm DC}^2}.
\end{equation}
According to (\ref{sy}), (\ref{ey2}) and (\ref{ey}), the variance of the clipping distortion noise, $\sigma_{\rm c}^2$, can be calculated by:
\begin{equation}
\begin{aligned}
&\sigma_{\rm c}^2 = {\rm E}\left[Y_{\rm c}^2\right] - {\rm E}^2[Y_{\rm c}]\\
&= \alpha_{\rm c}(1-\alpha_{\rm c})(\sigma_{\rm m-DCO}^2+B_{\rm DC}^2) 
- \bigg[B_{\rm DC}Q\left(-\frac{B_{\rm DC}}{\sigma_{\rm m-DCO}}\right)\\
&+ \sigma_{\rm m-DCO}\phi\left(\frac{B_{\rm DC}}{\sigma_{\rm m-DCO}}\right) - \alpha_{\rm c}B_{\rm DC} \bigg]^2.
\end{aligned}
\end{equation}
After DC biasing and clipping in time domain, the mean value of DCO-OFDM symbols is:
\begin{equation}
\begin{aligned}
{\rm E}&[\textbf{x}_{\rm{clipped}}(k)] = \int_{0}^{\infty} \frac{x}{\sigma_{\rm m-DCO}}\phi\left(\frac{x-B_{\rm DC}}{\sigma_{\rm m-DCO}}\right){\rm d}x \\
& = B_{\rm DC}Q\left(-\frac{B_{\rm DC}}{\sigma_{\rm m-DCO}}\right) + \sigma_{\rm m-DCO}\phi\left(\frac{B_{\rm DC}}{\sigma_{\rm m-DCO}}\right).
\end{aligned}
\end{equation}
Thus, the normalized DCO-OFDM symbols can be described as a Gaussian distribution from 0 to $\infty$ with mean value and standard deviation:
\begin{equation}
\rho_{\rm DCO} = \frac{B_{\rm DC}}{{\rm E}[\textbf{x}_{\rm{clipped}}(k)]},
\end{equation}
\begin{equation}
\sigma_{\rm x-DCO} = \frac{\sigma_{\rm m-DCO}}{{\rm E}[\textbf{x}_{\rm{clipped}}(k)]}.
\end{equation}
Similar to PQ SPAD ACO-OFDM, the nonlinear gain factor of PQ SPAD DCO-OFDM is:
\begin{equation}
\alpha_{\rm PQ-DCO} = \left. \frac{{\rm E}\left[N(x)z_{\rm PQ}\Big(N(x)\Big)\right]}{{\rm E}\left[N(x)^2\right]} \right |_{\rho = \rho_{\rm DCO}, \sigma_{\rm x} = \sigma_{\rm x-DCO}}, \\
\label{aPQDCO}
\end{equation}
Then, the equations, (\ref{ey2}) and (\ref{ey}), become:
\begin{equation}
{\rm E}\left[Y^2\right] = {\rm E}\left[z_{\rm PQ}\Big(N(x)\Big)^2\right] - \alpha_{\rm PQ-DCO}^2{\rm E}\left[N(x)^2\right],
\end{equation}
\begin{equation}
{\rm E}[Y] = {\rm E}\left[z_{\rm PQ}\Big(N(x)\Big)\right] - \alpha_{\rm PQ-DCO}{\rm E}\left[N(x)\right],
\end{equation}
where $\rho = \rho_{\rm DCO}$ and $\sigma_{\rm x} = \sigma_{\rm x-DCO}$. As a result, $\sigma_{\rm Y-PQ-DCO}^2$ can also be achieved. As the clipping distortion effect is considered in DCO-OFDM, the final resulting SNR of the PQ SPAD DCO-OFDM system can be calculated by:
\begin{equation}
{\rm SNR}^{\rm PQ}_{\rm DCO} = \frac{G_{\rm DC}\alpha_{\rm c}^2\alpha_{\rm PQ-DCO}^2C_{\rm s}^2\sigma_{\rm x-DCO}^2}{R_{\rm DCO}(\alpha_{\rm c}^2\sigma_{\rm cp}^2 + \sigma_{\rm Y-PQ-DCO}^2 + \sigma_{\rm N-PQ}^2)}.
\end{equation}
where $R_{\rm DCO}$ is equal to $\frac{N-2}{2N}\log_2 (M)$ and $\sigma_{\rm cp}^2$ is the attenuated variance of $\sigma_{\rm c}^2$ due to the transmitter normalization and the photon counter. In PQ SPAD DCO-OFDM, $\sigma_{\rm cp}^2$ is given by:
\begin{equation}
\sigma_{\rm cp}^2 = \left(1-\frac{G_{\rm DC}}{R_{\rm DCO}}\right)\frac{C_{\rm s}^2\sigma_{\rm c}^2}{{\rm E}^2[\textbf{x}_{\rm{clipped}}(k)]}.
\end{equation}
Finally, the analytical BER performance of PQ SPAD DCO-OFDM can be achieved by the corresponding BER calculation function as used in (\ref{BERPA}).

\subsection{OFDM with AQ SPAD}

As noted, AQ SPAD has a different output function with PQ SPAD. According to (\ref{AQnon}) and (\ref{BT}), the nonlinear transformation function of AQ SPAD OFDM is:
\begin{equation}
z_{\rm AQ}\Big(N(x)\Big) = \frac{N(x)}{1+C_{\rm t}N(x)} = \alpha_{\rm AQ}N(x) +Y_{\rm AQ}.
\end{equation}
To obtain the analytical equations of AQ SPAD OFDM, $z_{\rm AQ}\left(N(x)\right)$ is substituted into (\ref{ENzPQ}), (\ref{Ez2PQ}) and (\ref{EzPQ}) where $z_{\rm PQ}\left(N(x)\right)$ is replaced. Then following the same steps as in the analysis of PQ SPAD OFDM, the final resulting SNRs of AQ SPAD OFDM can be derived as:
\begin{equation}
{\rm SNR}^{\rm AQ}_{\rm ACO} = \frac{\alpha_{\rm AQ-ACO}^2C_{\rm s}^2\sigma_{\rm x-ACO}^2}{2R_{\rm ACO}(\sigma_{\rm Y-AQ-ACO}^2 + \sigma_{\rm N-AQ}^2)},
\end{equation}
\begin{equation}
{\rm SNR}^{\rm AQ}_{\rm DCO} = \frac{G_{\rm DC}\alpha_{\rm c}^2\alpha_{\rm AQ-DCO}^2C_{\rm s}^2\sigma_{\rm x-DCO}^2}{R_{\rm DCO}(\alpha_{\rm c}^2\sigma_{\rm cp}^2 + \sigma_{\rm Y-AQ-DCO}^2 + \sigma_{\rm N-AQ}^2)}.
\end{equation}
It is worth noting that both Possion distribution (\ref{pois}) and the exact distribution (\ref{arraydis}) need to be considered in the calculations of the shot noise component, $\sigma_{\rm N-AQ}^2$. In the case of \textit{Poisson distribution}, the variance is equal to the mean value. By using the same expression in PQ SPAD (\ref{sNPQ}), the shot noise component in AQ SPAD OFDM is:
\begin{equation}
\sigma_{\rm N-AQ}^2 = {\rm E}\left[\sigma_{\rm N-AQ}^2(x)\right] = {\rm E}\left[z_{\rm AQ}\Big(N(x)\Big)\right].
\label{sNAQ}
\end{equation}
In the case of the \textit{exact distribution}, based on the exact variance of the AQ SPAD array output (\ref{vAQe}), the shot noise can be derived as:
\begin{equation}
\begin{aligned}
&\sigma_{\rm N-AQ}^2 = {\rm E}\left[\sigma_{\rm AQ}^2(k)\right] \\
&= {\rm E}\left\{ N_{\rm SPAD}\lambda_{\rm N}^3\left[\frac{N(x)}{N_{\rm SPAD}} + g_{\rm N}^2\lambda_{\rm N}(1+\frac{2}{3}g_{\rm N}+\frac{1}{6}g_{\rm N}^2)\right]\right\},
\end{aligned}
\end{equation}
where $\lambda_{\rm N} = \left(1+C_{\rm t}N(x)\right)^{-1}$ and $g_{\rm N} = C_{\rm t}N(x)$. The performance of Poisson distribution and the exact distribution in PQ and AQ SPAD OFDM systems will be compared in the next section.

\section{Results and Discussion}

\begin{table}[!t]
\renewcommand{\arraystretch}{1.5}
\caption{Simulation Parameters}
\centering
\begin{tabular}{|l|l|}\hline
The active area of each SPAD device & 50.3 $\rm \mu m ^2$ \\\hline
Total area of the SPAD array & 0.16 $\rm mm^2$  \\\hline
The FF of the SPAD array, $C_{\rm FF}$ & 32.2\%  \\\hline
The PDP of each SPAD device, $C_{\rm PDP}$ & 20\%  \\\hline 
The DCR of each SPAD device, $N_{\rm DCR}$ & 7.27 kHz  \\\hline 
The APP of each SPAD device, $P_{\rm AP}$ & 1\%  \\\hline 
The dead time of each SPAD device, $\tau_{\rm d}$ & 13.5 ns \\\hline
Number of SPADs in an array, $N_{\rm SPAD}$ & 1024 \\\hline
The wavelength of the received light, $w_{\rm L}$ & 450 nm \\\hline
\end{tabular} 
\label{SP}
\end{table}

In this section, the analytical BER performance of SPAD-based DCO-OFDM and ACO-OFDM with the nonlinear distortion are compared. Moreover, the maximum bit rates of each scheme are found, which are limited by the nonlinear distortion effect. In the simulation, an ideal LED is assumed to emit blue light with a wavelength distribution centred around 450~nm. For the ideal LED transmitter, the recharged time of the circuit and on/off time of LED can be neglected, thus rising/falling edges have negligible effects on transmitted samples in the time domain. Therefore, in this study, each digital OFDM symbol is converted to intensity signals without any distortions. In addition, optical signals are assumed to pass through a flat fading channel and in the absence of background light. As a consequence, the received signals are still non-distorted intensities with additional shot noises. Thus, in this study, the signals are assumed to be affected by the receiver shot noise, nonlinear distortion and clipping distortion (low bias level DCO-OFDM). In previous research~\cite{SPADOFDM}, $T_{\rm s} = 1 \ {\rm ms}$ and $T_{\rm s} = 1 \ {\rm \mu s}$ were simulated and compared. Thus, in this study, these two scenarios are continued to be considered and analysed. A PQ SPAD array and an AQ SPAD array are considered with the same parameters as in~\cite{SPADOFDM} and as given in Table~\ref{SP}.

\subsection{BER Performance Comparisons}

\begin{figure}[!t]
\begin{center}
\includegraphics[width=0.48\textwidth]{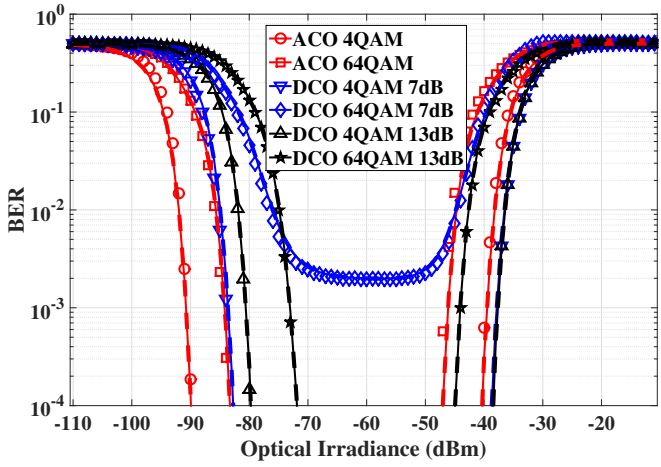}
\caption{BER performance of \textbf{PQ SPAD} ACO-OFDM and DCO-OFDM, $T_{\rm s} = 1 \ {\rm ms}$, simulation (symbols) vs. Poisson distribution theory (solid lines) vs. exact distribution theory (dashed lines).}
\label{PQ1ms}
\end{center}
\end{figure}

To present the BER performance and the receiver nonlinear distortion of SPAD-based OFDM, three definitions are given in this study. When the optical irradiance is larger than a threshold, the BER is below the target BER of $10^{-3}$. This threshold is defined as the minimum power requirement (MPR) of the system. When the optical irradiance increases and becomes larger than another threshold, the nonlinear distortion of SPAD receivers occurs, resulting in BER higher than $10^{-3}$. This threshold is defined as the maximum optical irradiance (MOI). The gap between the MPR and the MOI is defined as the low error area (LEA) where the system can maintain a low BER ($<10^{-3}$). For example, in Fig.~\ref{PQ1ms}, after the optical irradiance reaches $\textnormal{-}$~90.7~dBm, the BER of 4-QAM ACO-OFDM with PQ SPAD receivers is lower than $10^{-3}$, and after the optical irradiance reaches $\textnormal{-}$~39.6~dBm, the BER of the same scheme is higher than $10^{-3}$. Thus, the MPR of this scheme is $\textnormal{-}$~90.7~dBm; the MOI is $\textnormal{-}$~39.6~dBm; and the LEA is 51.1~dB. This means that 4-QAM PQ SPAD ACO-OFDM can be ideally used when the optical irradiance ranges from $\textnormal{-}$~90.7~dBm to $\textnormal{-}$~39.6~dBm.

\begin{figure}[!t]
\begin{center}
\includegraphics[width=0.48\textwidth]{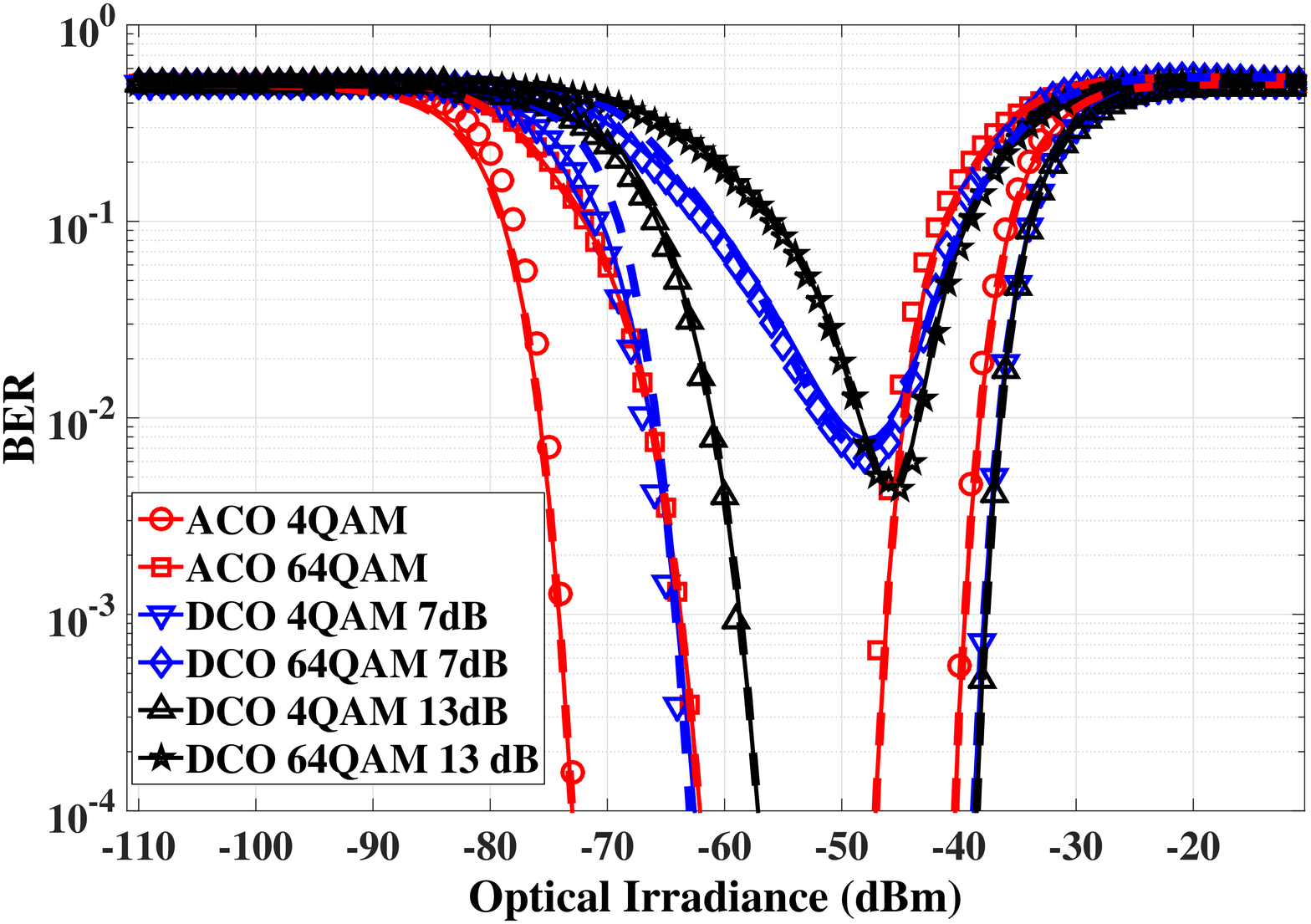}
\caption{BER performance of \textbf{PQ SPAD} ACO-OFDM and DCO-OFDM, $T_{\rm s} = 1 \ {\rm \mu s}$, simulation (symbols) vs. Poisson distribution theory (solid lines) vs. exact distribution theory (dashed lines).}
\label{PQ1us}
\end{center}
\end{figure}

\begin{figure}[!t]
\begin{center}
\includegraphics[width=0.48\textwidth]{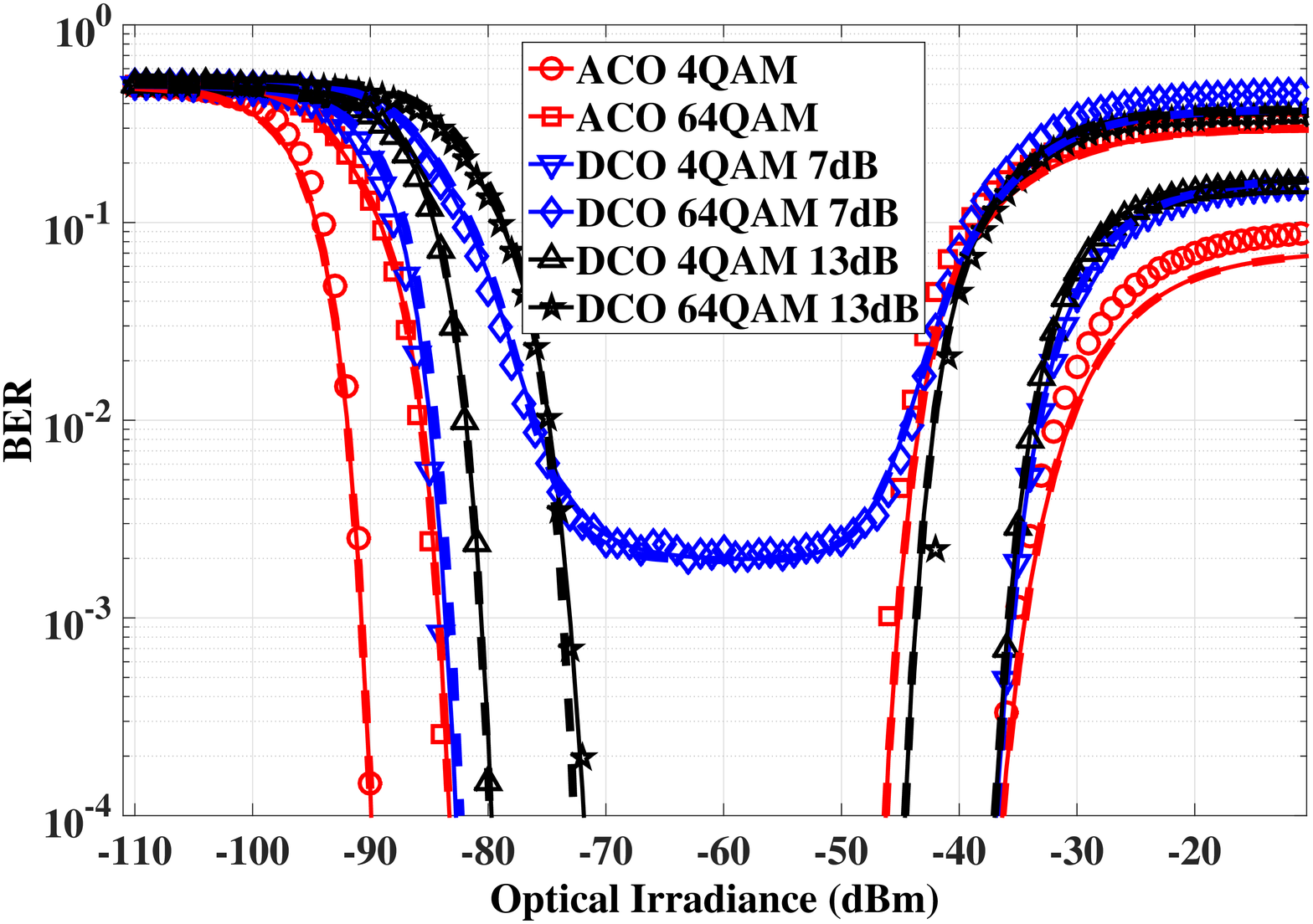}
\caption{BER performance of \textbf{AQ SPAD} ACO-OFDM and DCO-OFDM, $T_{\rm s} = 1 \ {\rm ms}$, simulation (symbols) vs. Poisson distribution theory (solid lines) vs. exact distribution theory (dashed lines).}
\label{AQ1ms}
\end{center}
\end{figure}

\begin{figure}[!t]
\begin{center}
\includegraphics[width=0.48\textwidth]{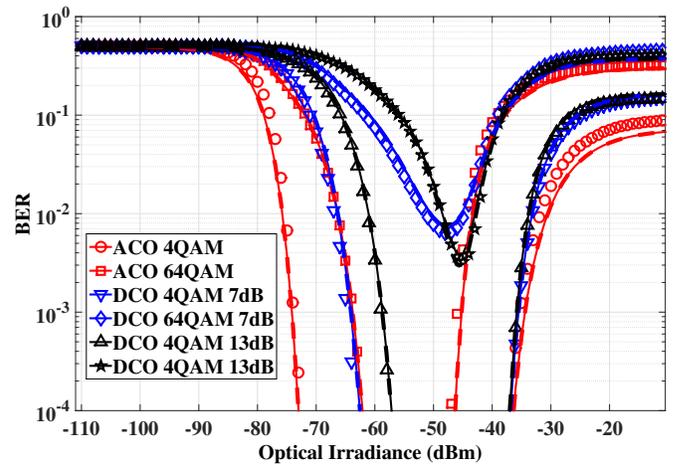}
\caption{BER performance of \textbf{AQ SPAD} ACO-OFDM and DCO-OFDM, $T_{\rm s} = 1 \ {\rm \mu s}$, simulation (symbols) vs. Poisson distribution theory (solid lines) vs. exact distribution theory (dashed lines).}
\label{AQ1us}
\end{center}
\end{figure}

The BER performance of the PQ SPAD ACO-OFDM and DCO-OFDM systems is presented in Fig.~\ref{PQ1ms} and Fig.~\ref{PQ1us} where $T_{\rm s} = 1 \ {\rm m s}$ and $T_{\rm s} = 1 \ {\rm \mu s}$. The analytical and simulation results confirm a very close match. It is shown that ACO-OFDM has a lower MPR than DCO-OFDM with the same constellation size. DC bias in DCO-OFDM consumes additional transmission power so that at the SPAD receiver side, the DCO-OFDM system receives more optical power than ACO-OFDM. As the additional nonlinear noise is doubled in ACO-OFDM~\cite{dsh2013non}, the nonlinear distortion in ACO-OFDM occurs earlier than DCO-OFDM. Thus it shows that the MOI of ACO-OFDM is lower than DCO-OFDM. On the whole, the LEA of ACO-OFDM is higher than DCO-OFDM. As a result, in SPAD-based OFDM systems, ACO-OFDM requires lower transmission power and has a longer operated interval when compared with DCO-OFDM. For different symbol periods, the schemes with a shorter symbol period ($T_{\rm s} = 1 \ {\rm \mu s}$) have higher MPRs than the methods with a longer symbol period ($T_{\rm s} = 1 \ {\rm m s}$). This means that with decreasing $T_{\rm s}$, the LEA reduces, which decreases the range of the received optical power. It is worth noting that for 64-QAM DCO-OFDM with 7~dB bias, the clipping distortion creates an error floor. Thus for higher constellation sizes, a higher DC bias may need to be applied in DCO-OFDM. However, the MPR increases with the additional DC bias, and at the same time, the LEA decreases. As shown in Fig.~\ref{PQ1us}, for 64-QAM DCO-OFDM with 13~dB bias, the MPR becomes higher than the MOI. Thus the BER of the system is always above $10^{-3}$ when $T_{\rm s} = 1 \ {\rm m s}$. Therefore, this symbol period is unacceptable for this scheme.

Fig.~\ref{AQ1ms} and Fig.~\ref{AQ1us} show the BER performance of the AQ SPAD ACO-OFDM and DCO-OFDM systems as a function of the optical irradiance when $T_s = 1 \ {\rm m s}$ and $T_s = 1 \ {\rm \mu s}$. Compared with the BER performance of PQ SPAD OFDM (Fig~\ref{PQ1ms} and Fig~\ref{PQ1us}), these two systems have the same BER performances at low optical irradiance (around MPR). This is because the PQ SPAD devices have the same performance of linearity as the AQ SPAD devices when the number of incoming photons is low (Fig.~5). However, as the maximum count rate of PQ SPAD is lower than AQ SPAD, the MOIs of PQ SPAD OFDM systems are lower than AQ SPAD OFDM systems. In addition, since the MPRs of each system are the same, the LEAs of PQ SPAD OFDM systems are also lower than AQ-based systems. As a consequence, the PQ SPAD OFDM system is more readily affected by the nonlinear distortion and has higher limitation of the optical irradiance.

As given in Section~\uppercase\expandafter{\romannumeral3}, exact distributions of the PQ and AQ SPAD array are considered in this study and compared with Poisson distribution. It is shown that the simulation results are well matched with both the Poisson distribution and the exact distribution. When the optical irradiance is low, Poisson has the same distribution as the exact one (Fig.~6(a) and Fig.~7(a)); and when the nonlinear distortion occurs, the variance of the nonlinear additional noise dominates the performance of the system and the shot noise has a negligible effect on the BER performance. Thus Poisson distribution shows the same performance as the exact distribution in the SPAD-based OFDM system. Although the exact distribution can describe the actual distribution of the SPAD arrays, the Poisson distribution is easier to implement in the simulation.

\begin{figure}[!t]
\begin{center}
\includegraphics[width=0.48\textwidth]{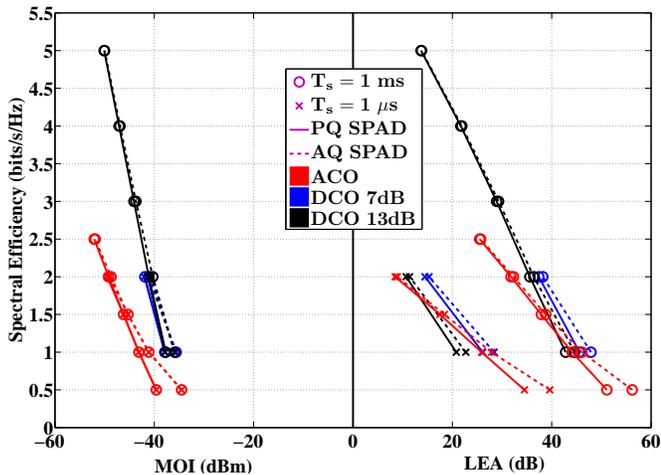}
\caption{The maximum optical irradiance (MOI) and low error area (LEA) of the SPAD-based OFDM systems; comparison between PQ SPAD (solid lines) and AQ SPAD (dashed lines) with ACO-OFDM and DCO-OFDM (7dB and 13dB DC bias); $T_{\rm s} = 1 \ {\rm m s}$ and $T_{\rm s} = 1 \ {\rm \mu s}$; BER = $10^{-3}$.}
\label{MOILEA}
\end{center}
\end{figure}

\subsection{Maximum Bit Rates}

Fig.~\ref{MOILEA} shows the MOI and LEA of the PQ SPAD and the AQ SPAD OFDM systems with different spectral efficiencies and symbol periods ($T_s = 1 \ {\rm m s}$ and $T_s = 1 \ {\rm \mu s}$). It is shown that the MOIs of all schemes decrease when the spectral efficiencies increase. High constellation size schemes have higher signal variances and peak-to-average power radio due to increasing of the probability of high intensity signals. Those high intensity signals are easily affected by the nonlinear distortion of SPAD receivers and increase the probability of error detections and demodulations. In addition, with the increase of constellation sizes, MPRs also increase. Therefore, with the increase of the spectral efficiency, LEAs of the systems rapidly decrease, as shown in Fig.~\ref{MOILEA}.

\begin{figure}[!t]
\begin{center}
\includegraphics[width=0.48\textwidth]{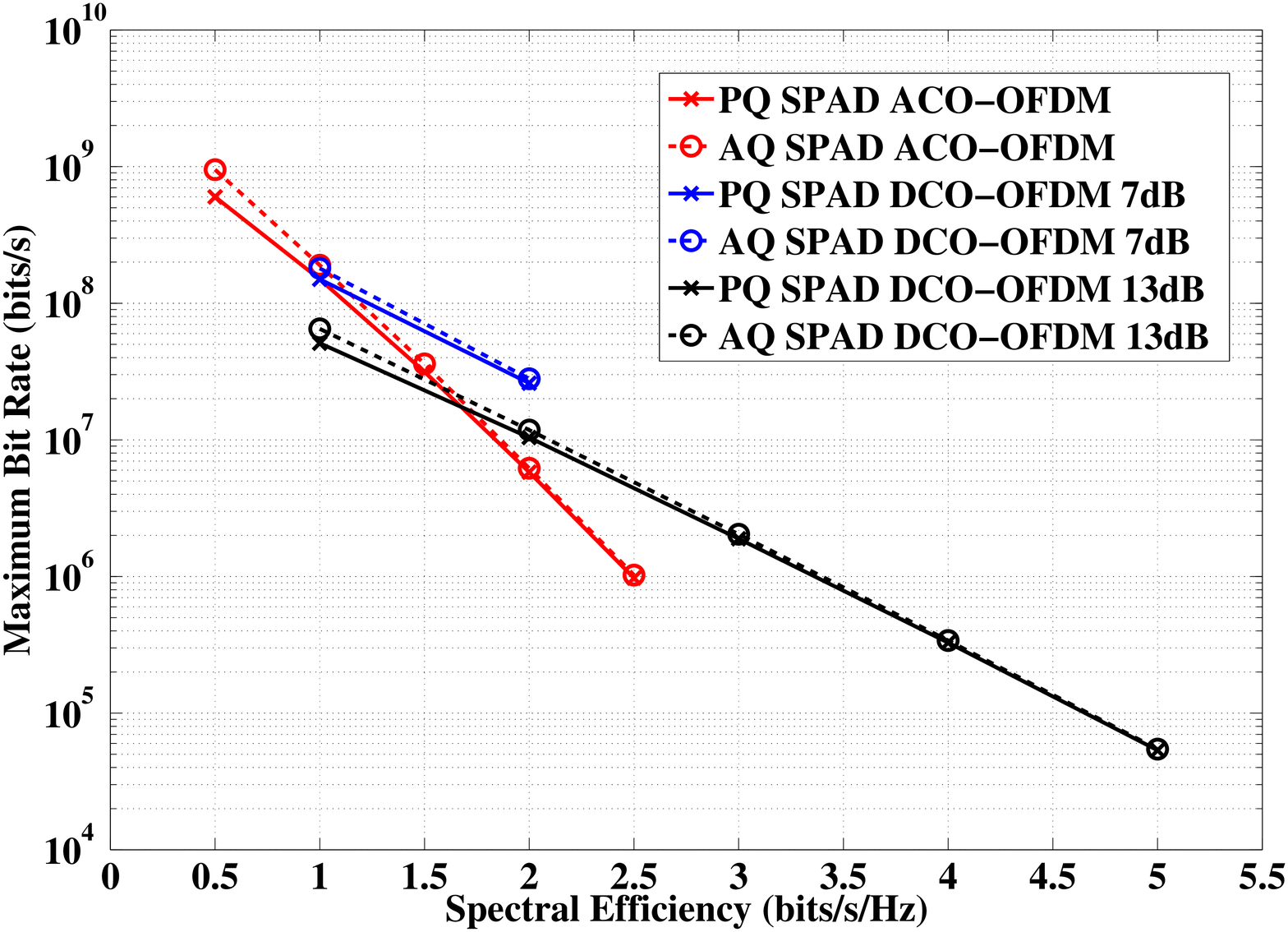}
\caption{Theoretical maximum bit rates of SPAD-based OFDM systems when BER = $10^{-3}$, PQ-SPAD (solid lines) vs. AQ-SPAD (dashed lines).}
\label{MAXBR}
\end{center}
\end{figure}

Note that the LEA of the SPAD-based OFDM system decreases when the symbol period becomes shorter. If the MOI is equal to the MPR, the BER of the corresponding SPAD-based system is always above $10^{-3}$. This means that the system cannot maintain a high-quality communication and thus the minimum symbol period can be obtained. As the bit rate is equal to the spectral efficiency divided by the symbol period, the maximum bit rate can also be obtained. By using the presented analytical BER model of the SPAD-based OFDM system in this study, the relationship between spectral efficiencies and theoretical maximum bit rates is obtained and is shown in Fig.~\ref{MAXBR}. The theoretical maximum bit rate of the SPAD-based OFDM system is up to 1~Gbits/s. Unlike the PD-based system, the increase of the spectral efficiency cannot bring a high bit rate in SPAD-based OFDM due to the limitation of the nonlinear distortion effect. However, since the SPAD receiver performs a significant enhancement on the power efficiency and sensitivity~\cite{SPADOFDM}, the maximum bit rate of the SPAD-based OFDM can be much higher than the conventional PD-based OFDM in the same transmission power condition.

\section{Conclusion}

In this paper, a complete analytical approach is presented for the performance analysis of the SPAD-based OFDM system with a receiver nonlinear distortion. The proposed theory shows very close agreement with the Monte Carlo simulation, thus confirming the validity of this analytical method. The presented analytical models provide an effective and accurate way to estimate system performance and to choose optimal parameters of the PQ and AQ SPAD receivers for the ACO-OFDM and DCO-OFDM system. For the assumed SPAD-based OFDM system, the nonlinear distortion has a significant effect on the BER performance when the optical irradiance is higher than $\textnormal{-}$~40 dBm. This maximum optical irradiance limits the maximum bit rate of the system, which is up to 1~Gbits/s, as shown in this study.

The SPAD receiver has a significantly enhanced sensitivity. This means that the SPAD-based OFDM system can be used in long distance transmissions, or it can be used in non-line-of-sight OWC links, in the uplink when illumination is not essential, or when lights are almost completely dimmed. However, due to such high sensitivity, an appropriate transmission power should be selected carefully so as to avoid the nonlinear distortion.

\bibliographystyle{IEEEtran}
\bibliography{./SPAD-OFDM-ANA}

\end{document}